\documentclass[floats,floatfix,amssymb,prd,twocolumn,nofootinbib,superscriptaddress]{revtex4-2}
\usepackage[utf8]{inputenc}
\usepackage{amsmath,amssymb}
\usepackage{amsfonts}
\usepackage{bm}
\usepackage{courier}
\usepackage{graphics,orcidlink}
\usepackage{float}
\usepackage{varioref}
\usepackage{mathtools}
\usepackage[normalem]{ulem} 
\usepackage{acronym}
\usepackage{float}
\usepackage[caption=false]{subfig}
\usepackage{lipsum}
\usepackage{multirow}
\usepackage{graphicx}
\usepackage{tensor}
\usepackage{verbatim}
\usepackage{wrapfig}
\usepackage{soul} 
\allowdisplaybreaks

\labelformat{section}{Section #1} 
\labelformat{subsubsection}{Section #1}
\labelformat{subsubsubsection}{Section #1}
\labelformat{equation}{Eq.~(#1)} 
\labelformat{figure}{Fig.~#1} 
\labelformat{subfigure}{Fig.~\thefigure#1} 
\labelformat{table}{Tab.~#1} 
\labelformat{appendix}{Appendix #1}

\begin{document}

\title{Probing the existence of a minimal length through compact binary inspiral}

\author{N. V. Krishnendu}
\email{k.naderivarium@bham.ac.uk}
\affiliation{Institute for Gravitational Wave Astronomy \& School of Physics and Astronomy, University of Birmingham, Birmingham, B15 2TT, United Kingdom}

\author{Aldo Perri}
\email{aldo.perri@studio.unibo.it}
\affiliation{Dipartimento di Fisica e Astronomia, Universit\`a di Bologna, Via Irnerio 46, 40126 Bologna, Italy}

\author{Sumanta Chakraborty}
\email{tpsc@iacs.res.in}
\affiliation{School of Physical Sciences, Indian Association for the Cultivation of Science, Kolkata-700032, India}

\author{Alessandro Pesci}
\email{pesci@bo.infn.it}
\affiliation{INFN Bologna, Via Irnerio 46, 40126 Bologna, Italy}

\begin{abstract}
Existence of a minimal length in spacetime geometries avoids several singular situations involving quantum theory and gravity. In this work, we show that the existence of such a minimal length also affects the gravitational wave (GW) waveform of any inspiraling binary black hole (BH) system by introducing a minimum frequency, below which the BHs behave as perfectly reflecting compact objects, while above they are identical to classical BHs. This leads to a significant imprint on the tidal heating term, appearing in the GW waveform at 2.5 post Newtonian order. Based on these modifications to the inspiraling waveform, it turns out that the detection of highly spinning and highly absorbing, almost classical BH like compact objects, inspiraling around each other, would be in tension with the quantum properties of BH geometries. The same would also be true if the zero point length exceeds the Planck length by a significant amount, suggesting that the zero point length, if it exists, must be of the same order as the Planck length, or smaller, purely from GW observations.
\end{abstract}

\maketitle
\section{Introduction}

Detection of gravitational waves (GWs) from the coalescence of binary black holes (BHs) by the LIGO-Virgo-KAGRA (LVK) detectors \cite{LIGOScientific:2016aoc, LIGOScientific:2019fpa, LIGOScientific:2020tif, LIGOScientific:2021sio, 
LVK:2022,
Nitz:2021zwj,LIGOScientific:2021usb,LIGOScientific:2020ibl,Olsen:2022pin} have enabled us to probe the strong gravity regime \cite{Will:2014kxa, Sathyaprakash:2019yqt, Berti:2016lat, Dreyer:2003bv, Berti:2018cxi, Berti:2018vdi, Barack:2018yly, Sathyaprakash:2009xs, Sathyaprakash:2019yqt}, along with possible existence of non-trivial effects (including quantum) at the horizon \cite{Agullo:2021, Cardoso:2019, Cardoso:2019rvt, Pani:2009ss} in an unprecedented manner. 
For quantum BHs, non-trivial quantum effects at the horizon scale are expected to modify the phasing of the emitted GWs during the inspiral of these BHs through tidal effects, namely tidal heating \cite{Poisson:2004cw, Alvi:2001mx, Chakraborty:2021gdf, Datta:2020rvo, Datta:2019epe, Chatziioannou:2012gq, Saketh:2022xjb} and tidal deformation \cite{Chakraborty:2024gcr, Chakraborty:2023zed, Cardoso:2017cfl, Creci:2021rkz, Chakravarti:2018vlt, Binnington:2009bb, Nagar:2011aa}, as well as through the existence of echoes in the final ringing phase \cite{Rosato:2025byu, Cardoso:2019, Abedi:2020ujo, Wang:2019rcf, Mark:2017dnq, Pani:2018flj, Konoplya:2011qq, Kokkotas:1999bd, Biswas:2022wah, Chakraborty:2022zlq}.
Of course, there will be modifications in the merger phase as well, but numerically solving the quantum corrected Einstein's equations for a binary BH system is beyond the scope of present day numerical techniques \cite{Scheel:2025jct, Paul:2024ujx}.

The fact that quantum effects at the horizon can be probed by GW observations may come as a surprise. 
In particular, the startling aspect from an experimental point of view is the result that even if the BH area changes by an amount as small as $\ell_{\rm p}^2$, where $\ell_{\rm p}$ is the Planck length, the GW frequencies involved in such a change in the area falls in the sensitivity range of the present GW detectors \cite{Agullo:2021}.
To see that the above is indeed the case, consider, for example, a Schwarzschild BH whose area changes by $\Delta A= \alpha \, \ell_{\rm p}^2$. 
This leads to the following change in the BH mass $\Delta M=(\Delta A/32\pi M)=(\alpha/32\pi M)\ell_{\rm p}^{2}$. 
The above change of mass produces GWs, following $\Delta M=\hbar \omega$, whose typical frequency for a solar mass BH reads $f_{\rm GW}\sim 1$kHz, precisely in the sensitivity band of the LVK GW detectors \cite{Agullo:2021}. 
Therefore, the discreteness of the area spectrum implies certain selection rules for the absorption of GW modes impinging on the BH horizon. 
Of course, the frequencies of GWs absorbed by the area quantized BH, depends on the choice of $\alpha$, which can be constrained using GW observations and in turn can rule out several phenomenological models of quantum gravity \cite{Datta:2021row, Chakravarti:2021jbv, Chakravarti:2021clm, Krishnendu:2025}.
For example, considering BHs as atoms of spacetime, Bekenstein had advocated $\alpha = 8\pi$ \cite{Bekenstein:1974}, which also arises from different grounds \cite{Maggiore:2008} and directly affects the Hawking quanta \cite{Chakraborty:2017opo, Lochan:2015bha}. Other proposed values for the constant $\alpha$ are $\alpha=4\ln 3$ \cite{Hod:1998}, or $\alpha=4\ln 2$ from BH microstate computation \cite{Mukhanov:1986, Barbero:2009}.

Besides the area quantization paradigm, which arises from many phenomenological as well as certain non-perturbative models of quantum gravity, e.g., loop quantum gravity \cite{Barbero:2009}, there can be several other alternatives. 
One such alternative, which naturally incorporates quantum effects into gravity is the zero point length paradigm \cite{Padmanabhan:1985jq, Padmanabhan:1985jdl, Garay:1994en, Nicolini:2022rlz}.
The main motivation is to regularize the singular coincidence limit of correlation functions of quantum fields on curved background \cite{Padmanabhan:1997}. 
In particular, it was realized that the zero point length corrected two-point function will have the following symmetry: $\sigma \to (\ell_{0}^{2}/\sigma)$, where $\sigma$ is the geodesic distance between two spacetime points and $\ell_0$ is the minimal length. 
The above symmetry transformation has also been related to the duality properties of the string theory \cite{Fontanini:2006, Bishop:2023hvz}. 
There have also been recent developments, where one uses the zero point length paradigm to construct an effective metric
\cite{KotE, KotF, KotI}, which has finite curvature invariants and modified Raychaudhuri equation leading to de-focussing of geodesics. 
Intriguingly, the Ricci scalar for the effective metric, in the coincidence limit, becomes $R_{\mu \nu}n^{\mu}n^{\nu}$,
with $n^\mu$ the vector tangent to the connecting geodesic, 
leading to a thermodynamic interpretation for gravitational theories \cite{KotF, KotI, Pesci:2020}. 
Thus it follows that the existence of a zero point length is a powerful tool to understand possible quantum effects of the spacetime on the smallest length scales 
\cite{Padmanabhan:2020, Padmanabhan:2016, ChaD}.
In summary, the existence of zero point length indeed helps in solving singularity problems in gravity theories through the introduction of quantum effects and endows them with a thermodynamic interpretation. 

In this paper, we wish to explore observational implications of the existence of a zero point length in the spacetime. In particular, we wish to explore the properties of the BH horizons arising out of the effective metric description from the existence of the minimum length $\ell_0$ \cite{KotE, KotF, KotI}. We expect the existence of a minimal length to introduce the notion of a minimal area (\emph{not} quantized area), which can lead to BHs harbouring non-trivial reflectivity. As we know, the existence of non-trivial reflectivity for ultra-compact objects are best probed using tidal properties of the compact objects, with the tidal heating being the most dominant one, as it appears in the GW waveform at 2.5 PN. Tidal heating is essentially the absorption of GWs by the BH horizon, leading to growth of the BH area and hence the growth of its mass~\cite{Chatziioannou:2012gq,Chatziioannou:2016kem,Hartle1973,Poisson2014}. For example, if the object is perfectly reflective, then it follows that the tidal heating will vanish identically, while for partially reflecting ultra-compact objects, the tidal heating will be non-zero, but smaller than that of the classical BH. This implies that the GW waveform will be different for a zero point length corrected BH, than classical BHs (CBHs). This is the feature that we wish to exploit in the present work in order to understand the implication of the existence of a zero point length on the GW waveform and also estimate the size of the zero point length compared to the Planck length.    

%

The paper is organized as follows: In \ref{minarea} we provide the derivation of the existence of a minimal area for a BH spacetime from the existence of a zero point length. Then in \ref{tidalheating} we discuss the implications of the minimal area on the GW waveform of inspiraling comparable mass BH binary through tidal heating, whose mismatch analysis has been performed in \ref{mismatch}. The waveform modelling for extreme mass ratio inspiral (EMRI) and the number of cycles computation has been discussed in \ref{EMRI}, following which we conclude in \ref{conc}. A heuristic derivation of the minimal area arising from minimal length has also been presented in \ref{app:Heuristic}.

\emph{Notations and Conventions:} We use mostly positive signature convention, with the flat Minkowski spacetime being expressed as $\eta_{\mu \nu}=\textrm{diag.}(-1,+1,+1,+1)$. All the four dimensional spacetime coordinates are denoted by Greek indices $\mu,\nu,\cdots$. We also set the fundamental constants $G$ and $c$ to unity.

\section{Minimal length to a minimal area}\label{minarea}

Aim of this section is to discuss how the existence of a minimal length can affect the area evolution of any surface and in particular of (the spatial cross-section of) black hole (BH) horizons.
To investigate this, namely the effects of a limiting length $\ell_{0}$ on the metric description of a spacetime, one may introduce the so-called minimum-length metric, or, quantum metric, or, qmetric for brief (see \cite{KotE, KotF, KotI} for details).
The qmetric incorporates the existence of a limit length through the modification of the geodesic interval $\sigma^2(p, P)$ (related to Synge's world function $\Omega=(\sigma^{2}/2)$) between two points $P$ and $p$ to a new spacetime interval $S=S(\sigma^2)$, such that $S(\sigma^2) \to \epsilon \ell_{0}^{2}$, when $p \to P$ along the geodesic connecting the two points, where $\epsilon = \pm 1$ with plus (minus) sign for space (time) separations (we use mostly-positive signature).

Due to the intrinsic non-locality present at the smallest scales, bi-tensors (i.e., tensorial quantities which depend on two points, like e.g., the Synge's world function itself, which is a bi-scalar) are taken as the basic building blocks of the construction.
%
The requirement of a finite limit for the new quadratic interval at the coincidence looks like quite an impossible task, since it necessarily demands for some metric which is diverging at any point of the spacetime.
With the qmetric, however, one tries to get a handle on this precisely through the use of bi-quantities. 
The metric tensor gets promoted to a metric bi-tensor which has to diverge in the coincidence limit $p\to P$, at any $P$.
However, thanks to the bi-structure, we can
still have a sensible and workable expression of the effective metric at any $p$ around each given fixed $P$.   
All the geometrical quantities depend, like the metric tensor, on the field point $p$, as well as the base point $P$.
Of course, some of them, as the metric tensor itself, might diverge in the coincidence limit $p \to P$, at any $P$, but some geometrical quantities of interest, most notably the Ricci scalar, do remain finite \cite{KotE}. 

This applies, almost unmodified, also to null separations \cite{QuantumMetricNull}, in spite of the quadratic interval vanishing identically in this case.
What happens in the case of null geodesics is our prime importance here, as the horizon generators of a BH are null. 
In the case of null geodesics, we consider a local observer $O$ at the base point $P$ and affinely parameterize any light ray through $P$ with its distance being described by the affine parameter $\lambda$. 
Thus, in the presence of a minimal length, one requires that the difference between the affine parameters between the two points $\lambda$, gets mapped to a new parameter $\widetilde\lambda = \widetilde\lambda(\lambda)$, which is affine according to the qmetric, such that $\widetilde\lambda(\lambda) \to \ell_{0}$ when $p \to P$ (i.e., when $\lambda \to 0$) along the null geodesic.
As it turns out, the above requirement on $\widetilde\lambda$, as imposed here for null geodesics, is the same as that on the modified geodesic separation $S$, presented above for spacelike/timelike geodesics, but, as in that case, it is not enough to fix completely the metric bi-tensor in terms of null geodesics. 
An additional requirement \cite{KotI, QuantumMetricNull}, which we do not delve into here, from causality, about the two-point functions in the two metrics (the ordinary metric and the qmetric) does the job.
We report here the expression for the metric bi-tensor for null separations (namely the case of specific interest here regarding BH horizon) \cite{QuantumMetricNull}:

\begin{align}
q_{\mu \nu}(p, P)&=a(p,P)\,g_{\mu \nu}(p)
\nonumber
\\
&+\left\{a(p,P)-1/\alpha(p,P)\right\}\,\left(l_{\mu}k_{\nu}+k_{\mu} l_{\nu}\right)_{p}~,
\end{align}
where, the metric $g_{\mu \nu}$, as well as the vectors $l_{\mu}$ and $k_{\nu}$ has been evaluated at the field point $p$. 
In addition, $l_{\mu}$ and $k_{\nu}$ are both null vectors, satisfying $\ell_{\mu}k^{\mu}=-1$, with $\alpha(p, P)$ and $a(p, P)$ being scalar functions of both $\widetilde{\lambda}$, and $\lambda$, which are given by,
\begin{align}
\alpha=\left(\frac{d\widetilde\lambda}{d\lambda}\right)^{-1}~;
\quad 
a=\frac{\widetilde\lambda^2}{\lambda^2}\,\Bigg(\frac{\Delta}{\Delta_{\widetilde\lambda}}\Bigg)^{\frac{2}{d-2}}~.
\end{align}
The above results hold for generic spacetime dimensions $d$, while for the purpose of this work we will concentrate on $d=4$. The above expressions for $\alpha$ and $a$, introduce the van Vleck determinant $\Delta$, which is defined as \cite{vVl, Mor, DeWA, DeWB},
\begin{eqnarray}
\Delta(p, P)=-\frac{1}{\sqrt{g(p) g(P)}} \,{\rm det} \Big[- \nabla_{\mu}^{(p)} \nabla_{\nu}^{(P)} \, \frac{1}{2} \sigma^2\Big]~.
\end{eqnarray}
The other bi-quantity $\Delta_{\widetilde\lambda}$ is the same as the one above, with $p$ taken at $\widetilde\lambda$ along the null geodesic.

An immediate property arising out of the above construction for the qmetric is the fact that ``area'' transverse to the geodesics connecting two nearby points tends to a finite value in the coincidence limit \cite{Pad06, QuantumMetricNull, ChaD}; this holds true, in particular for any two dimensional areas transverse to null geodesics in four spacetime dimensions.
In the present context, we are interested in understanding a similar result, i.e., if any change of area of a macroscopic surface also has a minimum cut off.
More precisely, if GWs with certain energy $E$ falls on a BH horizon, then the horizon may not absorb the GWs, unless the energy $E$ is larger than some threshold value, related to the zero point length $\ell_{0}$. 
In what follows, we will first determine this threshold value for BH area and then shall apply the result to determine the phase change in the GW signal emanating from an inspiralling binary BH system. 
This will be achieved by regarding the spacetime as a collection of coincidence events, providing an operational view.
%

For this purpose, we consider an area element of the BH horizon, as two dimensional spatial surface describing the spatial cross-section of a null geodesic congruence generating the horizon.
Note that classically the cross-sectional area of the horizon is irreducible, i.e., it can not decrease. 
From the point of view of an observer stationary in the spacetime metric and hovering just outside the horizon, the infalling matter causes the horizon to expand.  
As a consequence, a pencil of radially outgoing light rays, escaping very slowly from the BH, from a region outside the horizon, but very close to it, forms a slightly diverging (expansion scalar $\theta>0$) null congruence. After the lump of matter enters the horizon, these become new generators of the horizon, with final expansion being $\theta =0$. 
An equivalent picture exists from the point of view of a local free-falling observer 
as well. 



We consider the absorption of matter/energy by a generic stationary horizon and define the base point $P$, for the qmetric description, as the event corresponding to the crossing of the horizon by the matter/energy.
We choose the reference frame to be at rest with the matter, which is located at the spatial origin, and we assume that matter crosses the horizon along the $X$ direction with velocity $c$, i.e., from the perspective of the observer, the horizon comes to matter with velocity $c$ along $X$.
%
As null geodesic $\gamma$, we consider the light ray just outside the horizon, which moving along positive $X$ direction becomes a part of the horizon at $P$ as the matter/energy is engulfed by the BH.
We think of it as affinely parameterized with $\lambda=|X|$, and take as field point $p$, any spacetime event along the light ray trajectory $\gamma$, which reaches the matter/energy precisely at $P$. Note that $p$ is at $X, T<0$ and $\lambda=|X|=c|T|\equiv \lambda_p$.
The coincidence limit is obtained by taking $p(\lambda_p)$ approaching $P$.

Given the above setup, the zero-point area associated with the qmetric to the event $P$ can be computed as described below.
From \cite{QuantumMetricNull} it follows that the limiting area of the qmetric at $P$ is given by, 
\begin{eqnarray}\label{LimitArea}
dA_{0}=\lim_{\lambda_p\to 0} d\widetilde{A}=\frac{\ell_{0}^{2}d\Omega}{\Delta_{\ell_{0}}}\,,
\end{eqnarray}
where, we have used the fact that $l^{\mu} = (1, 1, 0, 0)$ is the tangent vector to the null geodesic $\gamma$
and, $\Delta_{\ell_{0}}$ is the van~Vleck determinant evaluated at the field point $p=\bar{p}$ with $\lambda_{\bar{p}}=\ell_{0}$ along $\gamma$, i.e., $\Delta_{\ell_{0}}=\Delta(\bar p, P)$. 
%
The above result follows from Eq. (42) of \cite{QuantumMetricNull}, when applied to four dimensional spacetime, and uses the fact that in the $\lambda_{p}\to 0$ limit, $\widetilde\lambda(\lambda_{p}) \to \ell_{0}$,
with $\tilde\lambda(\lambda_p)$ being the qmetric affine parameter corresponding to the given value $\lambda_p$ of the affine parameter of the ordinary metric.

In order to obtain the limiting area increment $\Delta A_0$ at $P$, we simply have to integrate over the full solid angle around the direction of approach of $p$ to $P$, i.e.,
\begin{eqnarray}
\Delta A_{0}=\lim_{\lambda \to 0} \int d\widetilde{A}=\ell_{0}^{2}\int_{\Omega=4\pi}\frac{d\Omega}{\Delta_{\ell_{0}}}\,.
\end{eqnarray}
In the approximation $\Delta_{\ell_{0}}\approx 1$ in any direction, which is generically satisfied as it requires $\ell_{0}\ll (1/\sqrt{\cal R})$, where $\cal R$ is a typical component of the Ricci tensor, we thus obtain
\begin{eqnarray}
\Delta A_{0}=4\pi \ell_{0}^{2}\,.
\end{eqnarray}
Therefore, the existence of a minimum length, also predicts a minimum area for the BH horizon. 
The residual area $\Delta A_{0}$ should be thought of as that of a spherical surface of radius $\ell_0$, which needs to be added to the BH horizon, leading to its minimal change due to infall of matter/energy through the horizon. If the area change is larger than $\Delta A_{0}$, then it will be consistent with the existence of minimal length, as the field point $p$ and the base point $P$ will have an affine distance between them, greater than $\ell_{0}$.   
All in all, the above analysis implies that, if some matter/energy falls on the BH horizon, and if the area change of the BH horizon due to this process is smaller than $\Delta A_{0}$, then that matter/energy will be reflected back from the horizon, and hence the quantum nature of the BH will be dominant in this regime. On the other hand, if the area change of the BH horizon is greater than $\Delta A_{0}$, then it will be absorbed and hence the BH will behave classically. This feature, as we will describe in the next section, will have direct implications for GW emission from compact binary inspiral. 

\section{Implications of minimal area for tidal heating}\label{tidalheating}

In this section, we will relate the existence of a minimal area, derived in the previous section, from the presence of a minimal length, with the gravitational wave phasing, through the tidal heating experienced by a system of binary BHs, inspiraling around each other. First of all, we notice that the presence of a minimal length, within the qmetric approach, demands the change in BH area to have a lower bound, $\Delta A_{\rm min}=4\pi \ell_{0}^{2}$. Using the laws of BH mechanics for a rotating BH of mass $M$ and angular momentum $J$, the above minimum area change can be translated to a minimum change of BH mass as, 
\begin{align}
\Delta M_{\rm min}=\left(\frac{\kappa}{8\pi}\right)4\pi \beta^{2} \ell_{\rm p}^{2}+\frac{\Delta J^{2}}{4M\left(M^{2}+\sqrt{M^{4}-J^{2}}\right)}~.
\end{align}
Here we have defined the parameter $\beta\equiv(\ell_{0}/\ell_{\rm p})$, where, $\ell_{\rm p}$ is the Planck length, and is a free parameter of the problem. Moreover, the angular momentum can also be considered to be quantized, such that $J^{2}=\hbar^{2}j(j+1)\approx \hbar^{2}j^{2}$, since the BH is a macroscopic system and hence $j\gg 1$. As our interest is in GWs, we choose $\Delta j=2$, since this corresponds to the lowest change of mass and gravitons are spin-2 particles\footnote{Technically speaking, if we are just interested in the fact that gravitons are spin-2 particles, then the final BH can have angular momentum from $(j+2)$ to $(j-2)$, and hence there can be situations where $\Delta j=-2$ as well, with certain probability given by the respective Clebsch-Gordan coefficients. However, in practice, we are only interested in the quadrupolar radiation, with $\ell=2=m$, and hence the magnetic part of the total angular momentum must reach till $(j+2)$, which suggests that we will only have contributions from $\Delta j=2$.}. Therefore, the minimum frequency, a BH within the zero point length paradigm, can absorb, is given by, 
\begin{align}\label{minfreq}
\omega_{\rm min}=\frac{\Delta M_{\rm min}}{\hbar}=\left(\frac{\kappa \beta^{2}}{2}\right)+2\Omega_{\rm h}~,
\end{align}
where, the surface gravity $\kappa$ and the angular velocity $\Omega_{\rm h}$ of the BH horizon are given by,
\begin{align}
M\kappa=\frac{\sqrt{1-\chi^{2}}}{2\left[1+\sqrt{1-\chi^{2}}\right]}~,
\,
M\Omega_{\rm h}=\frac{\chi}{2\left[1+\sqrt{1-\chi^{2}}\right]}~,
\end{align}
with $\chi\equiv(J/M^{2})$ being the dimensionless angular momentum. The consequence of \ref{minfreq} being, if a GW with frequency $\omega<\omega_{\rm min}$ falls on the BH it will be perfectly reflected, while for $\omega>\omega_{\rm min}$ it will be absorbed. This suggests the following reflectivity for a BH with minimal length: 
\begin{align}
|\mathcal{R}(f)|^{2}&=1 \qquad f<f_{\rm min}
\\
&=0 \qquad f\geq f_{\rm min}~,
\end{align}
where, $f=(\omega/2\pi)$. Since the effective potential experienced by perturbations vanish near the BH horizon, it follows that near the BH horizon, such perturbations can be expressed as $e^{-i\bar{\omega} r_{*}}+\mathcal{R}(\omega)e^{i\bar{\omega}r_{*}}$, and hence non-zero reflectivity implies that there will be outgoing perturbation modes near the horizon. Here, $\bar{\omega}\equiv \omega-m\Omega_{\rm h}$, is the GW frequency in the co-rotating frame of reference. As the above relation implies, for $\omega \geq \omega_{\rm min}$, the reflectivity vanishes, and hence the perturbation modes are purely ingoing at the horizon, typical of a classical BH. For $\omega<\omega_{\rm min}$, on the other hand, the reflectivity is unity (modulo a phase factor), and hence the BH becomes perfectly reflective, implying that the amplitude of ingoing perturbation mode is same as that of the outgoing perturbation mode. Note that, in a binary system, the minimum frequency $f_{\rm min}$ will be different for the individual BHs, and hence one need to introduce two such reflective functions $\mathcal{R}_{1}(f)$ and $\mathcal{R}_{2}(f)$ for both the BHs individually. We would like to point out that for advanced LIGO GW detectors, there is a threshold of 20 Hz, and hence it follows that for $f_{\rm min}\leq 20$ Hz, the quantum BH (or, equivalently zero-point length corrected BH, QBH) will behave as a classical BH for all intents and purposes. The existence of a non-trivial frequency dependent reflectivity will be the main discriminating feature between classical and quantum BHs, which will play the dominant role in distinguishing them using the effect of tidal heating on the phasing of GW waveform. 

\begin{figure}[t]
\includegraphics[width=0.5\textwidth]{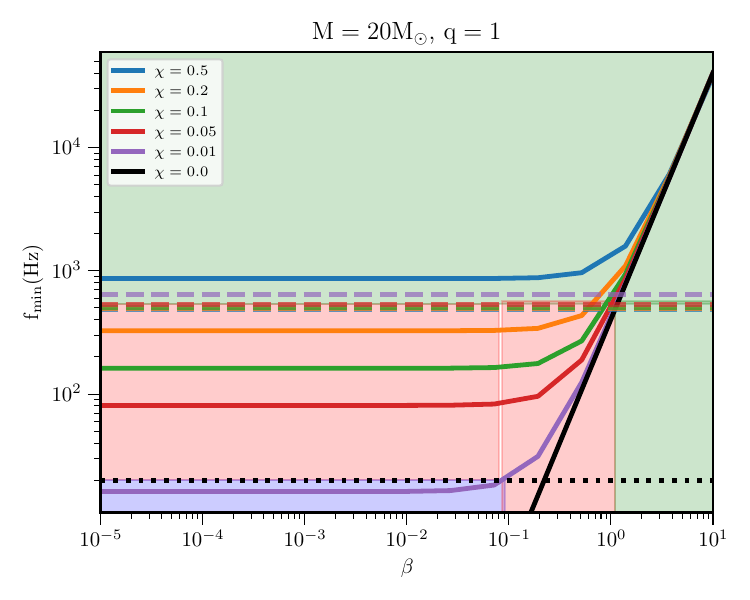}
\caption{The value of $f_{\rm min}$ in Hz has been plotted against the parameter $\beta$, associated with zero-point length, for a binary BH system having fixed total mass $20M_{\odot}$ and with mass ratio $q=1$. We assume that both the BHs in the binary system have the same spin (so that, both the quantum BHs have the same $f_{\rm min}$) and we plot the $f_{\rm min}$ for various choices of the dimensionless spin, from $\chi=0.0$ (black) to $\chi=0.5$ (blue). We have also depicted the zero spin case. The black dotted line at the bottom of the plot depicts the 20 Hz cut off of the Advanced LIGO type detectors, while the dashed lines around 500 Hz, depicts the ISCO frequencies. The plot has been divided into three regions: (a) perfectly absorbing (light purple region at bottom), (b) partially reflecting (light red region in the middle) and (c) perfectly reflecting (light green region at the top and right most part of the plot). See \ref{tab1} for more details.}
\label{fig:f_min_alpha_spins}
\end{figure}

\begin{table*}[htbp]
\begin{tabular}{|c|c|c|}
\hline
   Perfectly absorbing   &  Partially absorbing  & Perfectly reflecting  \\
    $(\mathcal{R}=0)$  &   $(0<\mathcal{R}<1)$ &  $(\mathcal{R}=1)$\\
   \hline
   \hline
   (i) Small spin ($\chi\lesssim 0.02$) & (i) Small and intermediate  & (i) High spin ($\chi\gtrsim 0.25$) \\
    and              & spin ($\chi\lesssim 0.25$) and & and any values of $\beta$ \\
    small $\beta$ ($\beta \lesssim 0.1$) & intermediate $\beta$ (0.1$\lesssim \beta \lesssim $2) & (ii) High $\beta$ ($\beta\gtrsim2$) \\
        &  (ii) Intermediate spin  & and arbitrary spin  \\
        & ($0.02\lesssim\chi\lesssim0.25$) and  & \\
        &  small $\beta$ ($\beta\lesssim 0.1$)  &\\
   \hline
   \hline
\end{tabular}
\caption{We present the summary of the relation between the reflectivity of the zero point length corrected BH with the $(\chi,\beta)$ parameter space. As evident, all the three cases are present, namely perfectly absorbing, partially reflecting as well as perfectly reflecting, depending on various choices for $\beta$ and $\chi$. In particular, small $\beta$ and small rotation is diametrically opposite to high $\beta$, or, high spin.}
\label{tab1}
\end{table*}

\begin{figure}[tbp]
\includegraphics[width=0.5\textwidth]{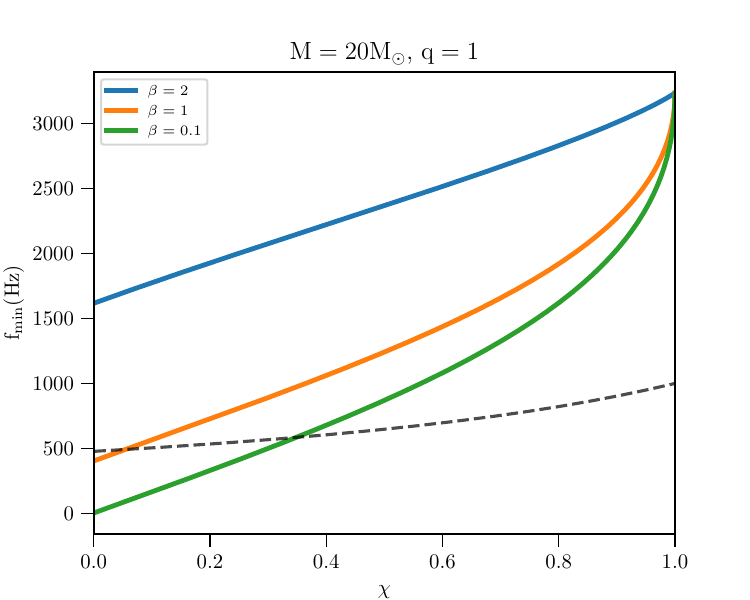}
\caption{Here we have plotted the minimum frequency $f_{\rm min}$ as a function of the dimensionless rotation parameter $\chi$ for an identical BH binary system with fixed total mass $20M_{\odot}$ and mass ratio $q=1$, for different $\beta$ values, namely 0.1 (green),  1 (orange) and 2 (blue). The frequency of the GW associated with the ISCO has been plotted against the rotation parameter, and depicted by the black dotted curve, for completeness.} 
\label{fig:f_min_beta_spin2d}
\end{figure}

The behaviour of the minimum frequency $f_{\rm min}$ with $\beta$, for an identical BH binary system (mass ratio $q=1$) with individual masses being few tens of solar masses, but for different choices of the BH rotation parameter has been presented in \ref{fig:f_min_alpha_spins}. As evident, for small rotation ($\chi\leq 0.02$) and small $\beta$ ($\beta<0.1$), the minimum frequency $f_{\rm min}$ lies below the 20 Hz cut off frequency, and hence the classical and quantum BH will behave in an identical manner, as far as the advanced LIGO type GW detectors is concerned. On the contrary, for large rotation parameter ($\chi\gtrsim0.25$), irrespective of the choice of $\beta$, it follows from \ref{fig:f_min_alpha_spins} that the minimum frequency is above the inner-most stable circular orbit (ISCO) frequency and hence the quantum BH is maximally different from a classical BH and behaves as close to a point particle as possible. While, for intermediate rotation parameter ($0.02\lesssim \chi \lesssim 0.25$), and small or, intermediate $\beta$ ($\beta \lesssim 2$), it follows that the quantum BH does behave as a classical BH over a range of frequencies and is the domain where, we expect to obtain reasonable constraints on the parameter $\beta$ from GW observations. The above features can also be seen from \ref{fig:f_min_beta_spin2d}. Moreover, for $\beta\gtrsim 2$, as is evident from both \ref{fig:f_min_alpha_spins} and \ref{fig:f_min_beta_spin2d}, the quantum BH is totally distinct from classical BH, as it remains totally reflective and this fact is independent of the spin of the BH, as far as tidal effects in the GW waveform in the advanced LIGO type detectors are concerned.  

\begin{figure}[tbp]
\includegraphics[width = 0.5\textwidth]{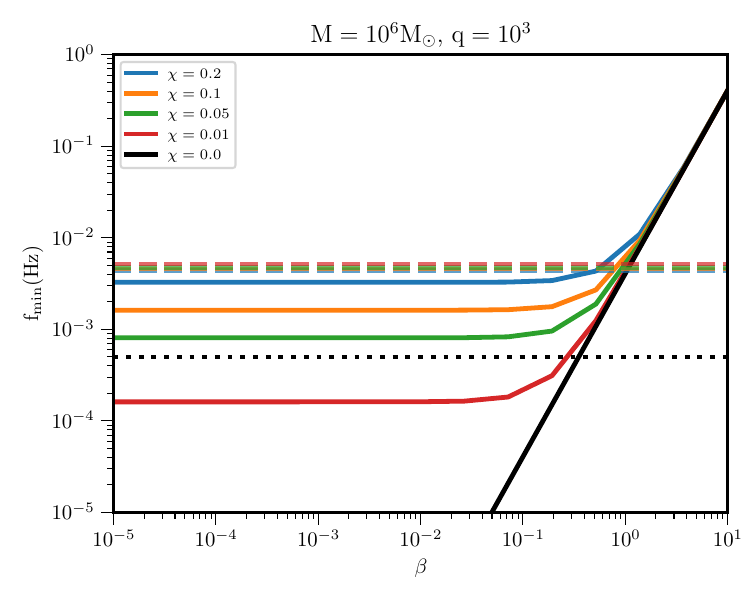}
\caption{The minimum frequency $f_{\rm min}$ in Hz has been plotted as a function of $\beta$ for an EMRI binary of total mass $10^{6}\rm{M_{\odot}}$ and a range of dimensionless spins, from 0.0 (black) to 0.2 (blue). The zero spin case has also been included. The dotted black line represents the lower cut-off frequency of $0.5$ mHz, associated with LISA, and the dashed lines represent the frequencies corresponding to the ISCO frequency of the Kerr BHs with varied spins, assuming a mass asymmetry of $10^{3}$.}
\label{fig:f_min_alpha_spins_EMRIN}
\end{figure}

It follows that the above conditions hold for supermassive BHs as well, with the minimum frequency $f_{\rm min}$ being scaled by the mass of the supermassive BH, and typically is in the mHz regime, or, lower. In which case, using the lower cut off frequency for LISA (which is $\sim 0.5$ mHz), it follows that (see \ref{fig:f_min_alpha_spins_EMRIN}) for large rotation ($\chi\gtrsim 0.2$) and large $\beta$ ($\beta>2$), the minimum frequency is always greater than the ISCO frequency, and hence the quantum BHs are purely reflective for all frequency ranges of interest. While for other rotation parameters of the BHs and for other choices of $\beta$, the quantum BH is partially or fully absorptive. The summary of the reflective nature of the zero point length incorporated BH with the $(\beta,\chi)$ phase space has been presented in \ref{tab1}. Finally, the behaviour of the minimum frequency at large $\beta$ is independent of the BH spin for smaller spins, but is different as the spin of the BHs increases, as evident from both \ref{fig:f_min_alpha_spins} and \ref{fig:f_min_alpha_spins_EMRIN}. In particular, in the extremal limit, $\kappa\to 0$, and hence the minimum frequency becomes independent of $\beta$, depending on the spin of the BH alone.

This suggests that if we detect the inspiral of either a solar mass BH using LVK or, next generation GW detectors, or, an EMRI using LISA, with very high spin, then the presence of the quantum nature of the BH, in particular, the quantization of angular momentum, argues that the QBH will be perfectly reflecting upto ISCO (independently of $\beta$). Therefore, if the matching of the detected GW waveform of this high spin compact binary inspiral with the theoretical GW waveform, assuming classical BH, reveals vanishingly small reflectivity, then it suggests that there is no angular momentum quantization, which will be at tension with quantum nature of gravity. Further, for slowly spinning BHs, $\beta\gtrsim2$ can be ruled out using next generation GW detectors, in particular from EMRI, as we will demonstrate later. This is because, for $\beta\gtrsim2$, the QBH is perfectly reflecting upto ISCO (for all spins). Therefore, if the detected GW waveform is inconsistent with large reflection from BHs, it suggests that $\beta\lesssim2$. Such an observation will rule out any model, in which the zero point length is larger than the Planck length, since the zero point length should be universal to all BHs. Thus, it seems that if the future GW detectors do not observe highly reflective BHs, then the zero point length must be constrained by the Planck length as an upper bound. As we will demonstrate, and it will also be clear from subsequent discussion that only for small spin and $0.1\lesssim\beta\lesssim2$, with high mass ratio, is the golden place to constrain $\beta$, as for smaller values of $\beta$, the reflectivity is governed by the spin alone.


\begin{figure*}[]
\includegraphics[width=0.3\textwidth]{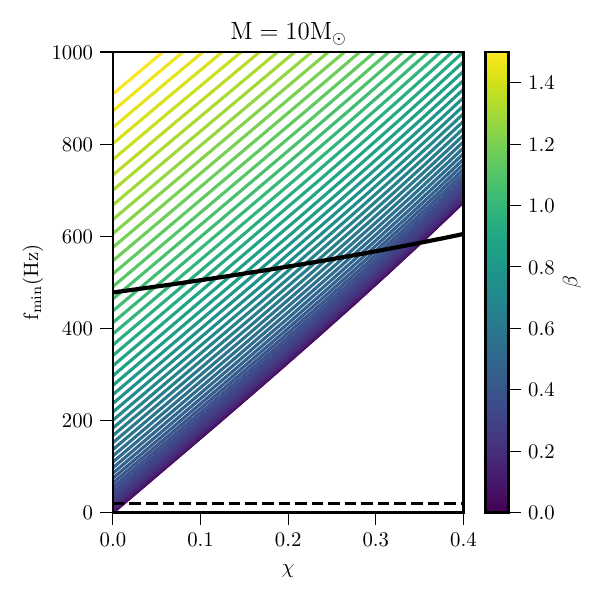}
\includegraphics[width=0.3\textwidth]{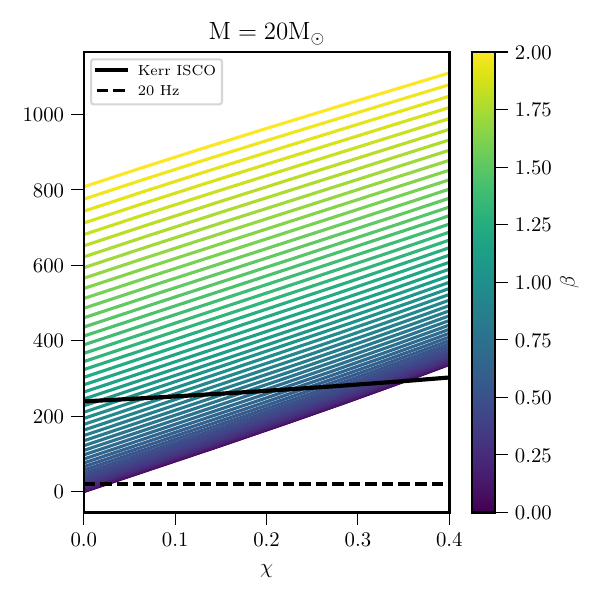}
\includegraphics[width=0.3\textwidth]{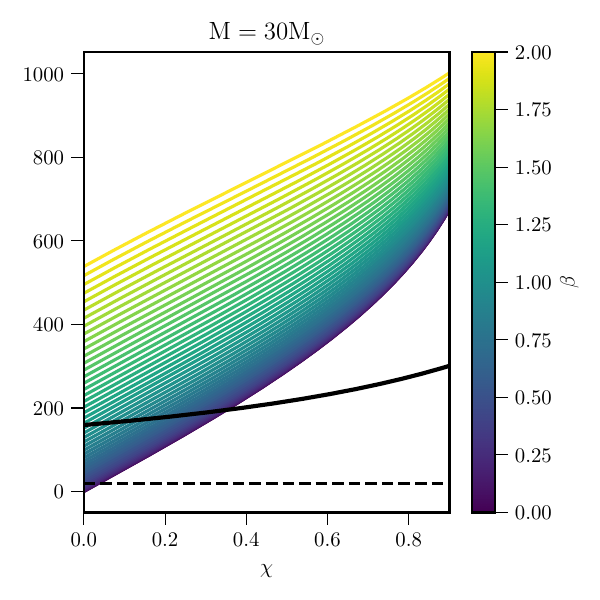}
\caption{The minimum frequency $f_{\rm min}$, beyond which there is no distinction between the classical and the zero point length corrected BH, has been shown for three different choices of the total mass of the binary, $10M_{\odot}$, $20M_{\odot}$, and $30M_{\odot}$ for an identical BH binary. The plots depict the variation of the minimum frequency $f_{\rm min}$ as a function of dimensionless rotation parameter $\chi$, with the colour bar showing the $\beta$ values.} 
\label{fig:f_min_beta_spin3d}
\end{figure*}

\begin{figure*}[]
\includegraphics[width = 0.3\textwidth]{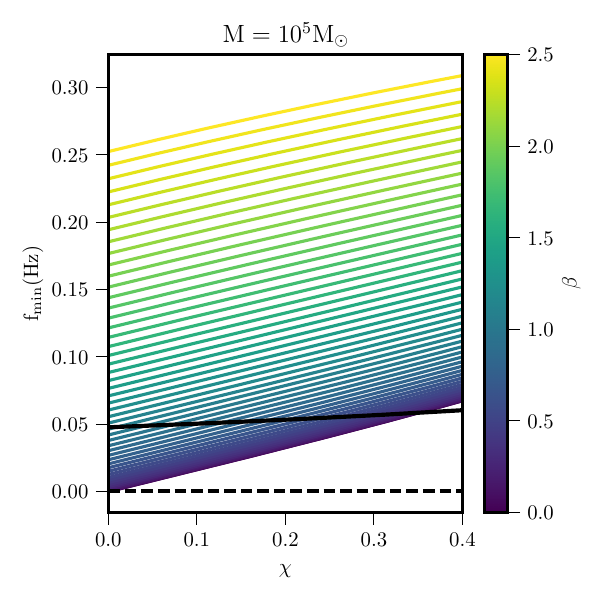}
\includegraphics[width = 0.3\textwidth]{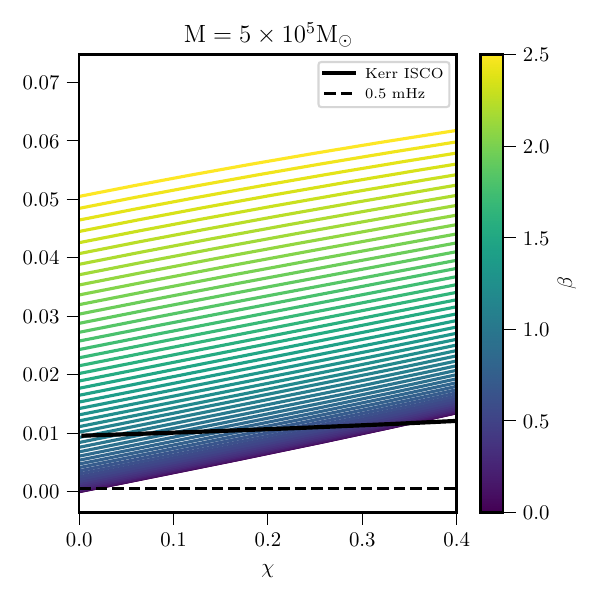}
\includegraphics[width = 0.3\textwidth]{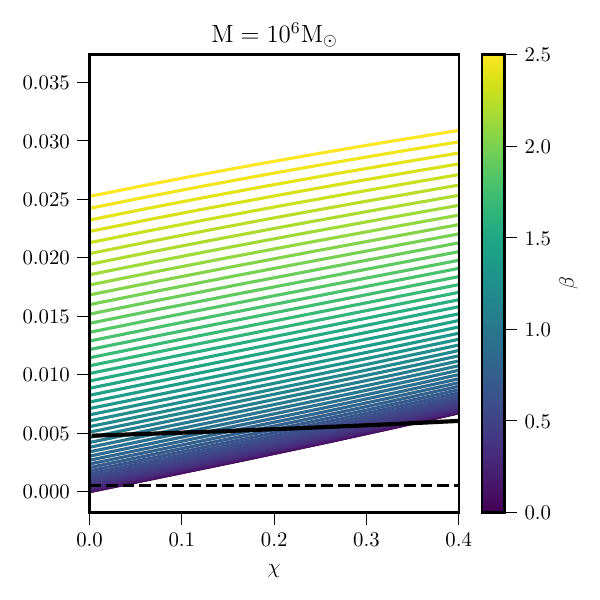}
\caption{The minimum frequency $f_{\rm min}$ in Hz has been plotted as a function of the dimensionless spin parameter $\chi$ for various $\beta$ values, as indicated in the adjacent colour bar, for three BH binaries of total masses $10^5 \rm{M_{\odot}}$, $5\times 10^5 \rm{M_{\odot}}$ and $10^6 \rm{M_{\odot}}$. The cut off frequency of $0.5$ mHz and the ISCO frequencies for these BHs have been indicated by dashed and solid black lines. The lower cut-off frequency is determined by the minimum sensitivity of LISA. As evident for smaller $\beta$, the $f_{\rm min}$ values are below the cut off frequency of LISA and hence behaves as a classical BH for all intents and purposes, while for larger values of $\beta$, the $f_{\rm min}$ values are above the ISCO frequencies and hence these BHs are perfectly reflecting.}
\label{fig:f_min_alpha_spins_EMRI}
\end{figure*}

The previous discussion depicts the influence of the rotation and the parameter $\beta$ associated with quantum BH on the minimum frequency of GW absorption, $f_{\rm min}$. The effect of the mass and spin together on the minimum frequency, for an identical BH binary system, for different $\beta$ has been depicted in \ref{fig:f_min_beta_spin3d}. The figure demonstrates that for smaller $\beta$, dependence of $f_{\rm min}$ on rotation and mass is different than the one for larger values of $\beta$. This is because, as evident from \ref{minfreq}, for small $\beta$, the minimum frequency is governed by $\Omega_{\rm h}$, while for large $\beta$, both $\kappa$ and $\Omega_{\rm h}$ contributes. 
Moreover, it is clear from \ref{fig:f_min_beta_spin3d} that, irrespective of the choice of $\beta$, BH binaries with larger spin and larger mass effectively behave as perfectly reflecting objects, in complete contrast to classical BHs. These features can also be observed in \ref{fig:f_min_beta_spin3d}, which shows that quantum BHs with smaller $\beta$ and rotation is effectively a classical BH, while larger $\beta$, keeps the quantum nature of BH even upto the ISCO frequency (shown by the black curve in \ref{fig:f_min_beta_spin3d}). The same holds true for EMRIs as well, where also larger spin and larger mass makes the BH perfectly reflective, making the distinction of quantum BH from classical BH a prominent one (see \ref{fig:f_min_alpha_spins_EMRI}). While for smaller $\beta$, the minimum frequency is below the cutoff frequency for LISA for small rotation and hence there is no difference between quantum and classical BH, while for larger rotation parameters, the quantum BH becomes partially reflecting. This is independent of the total mass of the binary, which is also evident from all the plots in \ref{fig:f_min_alpha_spins_EMRI}. 

Having described the implications of the existence of a minimum frequency and its correlation with BH parameters, we would like to discuss its effect on the GW waveform, which in turn requires determination of the GW flux. The GW flux emitted in the centre of mass frame will be distributed into a part reaching infinity and another part reaching the horizon. The GW flux entering the BH horizon is referred to as the tidal heating, and this corresponds to the following rate of change of BH mass within the zero point length paradigm, which effectively introduces an extra multiplicative factor $\mathcal{H}(f)\equiv 1-|\mathcal{R}(f)|^{2}$, referred to as the transmitivity of the BH \cite{Chakraborty:2021gdf,Datta:2020rvo,Datta:2019epe,Alvi:2001mx,Krishnendu:2024jkj}, 
\begin{align}
\dfrac{dM_{i}}{dt}&=\mathcal{H}_{i}(v)\left(\dfrac{dE}{dt}\right)_{\rm N}\left(\frac{M_{i}}{M}\right)^{3}\frac{v^{5}}{4}\Bigg\{-\chi_{i}\left(\hat{\boldsymbol{L}}_{\rm orb}\cdot \hat{\boldsymbol{J}_{i}} \right)
\nonumber
\\
&\qquad \qquad+\left(\frac{2r_{+i}}{M}\right)v^{3}\Bigg\}\left(1+3\chi_{i}^{2}\right)~.
\end{align}
Here, $i=1,2$ describes the two BHs in the binary, $M\equiv M_{1}+M_{2}$ is the total mass, $\chi\equiv (J/M^{2})$ is the dimensionless angular momentum, $r_{+}$ is the location of the event horizon, $\boldsymbol{L}_{\rm orb}$ is the orbital angular momentum, and the relative velocity in the centre of mass frame is given by $v=(\pi M f)^{1/3}$. The term $(dE/dt)_{\rm N}=(32/5)\eta^{2}v^{10}$, corresponds to the leading order energy loss, namely the quadrupolar radiation, where, $\eta \equiv (M_{1}M_{2}/M^{2})$. Therefore, the total flux through the horizon is given by, 
\begin{align}\label{flux_hor}
\mathcal{F}_{\rm H}=\sum_{i=1}^{2}\left(\dfrac{dM_{i}}{dt}\right)
=\left(\dfrac{dE}{dt}\right)_{\rm N}\left[-\Psi_{5}\frac{v^{5}}{4}+\Psi_{8}\frac{v^{8}}{2}\right]~,
\end{align}
where, the quantities $\Psi_{5}$ and $\Psi_{8}$ are defined as,
\begin{align}\label{Psi58}
\Psi_{5}&\equiv \sum_{i=1}^{2}\mathcal{H}_{i}(v)\Psi_{5(i)}^{\rm (BH)}~;
\quad 
\Psi_{8}&\equiv \sum_{i=1}^{2}\mathcal{H}_{i}(v)\Psi_{8(i)}^{\rm (BH)}~,
\end{align}
with $\Psi_{5(i)}^{\rm (BH)}$ and $\Psi_{8(i)}^{\rm (BH)}$ given by, 
\begin{align}\label{Psi5bh}
\Psi_{5(i)}^{\rm (BH)}&\equiv \left(\frac{M_{i}}{M}\right)^{3}\chi_{i}\left(\hat{\boldsymbol{L}}_{\rm orb}\cdot \hat{\boldsymbol{J}_{i}} \right)\left(1+3\chi_{i}^{2}\right)~,
\\
\Psi_{8(i)}^{\rm (BH)}&\equiv \left(\frac{M_{i}}{M}\right)^{4}\left(1+3\chi_{i}^{2}\right)\left(1+\sqrt{1-\chi_{i}^{2}}\right)~.
\label{Psi8bh}
\end{align}
Thus the flux due to tidal heating, i.e., absorption of GWs by the BH horizon depends on the existence of the zero point length, through the absorption terms $\mathcal{H}_{i}(f)$ for individual BHs. In particular, these absorption terms depend on the free parameter $\beta$ arising from the qmetric approach to the zero point length paradigm, and hence one expects to constrain this parameter using GW observations. Note that if the zero point length is identical to the Planck length, then $\beta=1$, however, the requirement of the cosmological information being $4\pi$ \cite{Padmanabhan:2017qvh} yields, $\beta=10^{4}$. This suggests that $\beta$ has a large spectrum and thus it would be interesting to see which values of $\beta$ is preferred by the GW observations. In particular, if $\beta\sim 10^{4}$ is preferred then it would suggest that the effect of zero point length at the early universe and the zero point length contributing to BH physics are intrinsically tied with one another. Unfortunately, as our previous discussion demonstrates, any value of $\beta\gtrsim 1$ will lead to perfectly reflecting BHs, which is seemingly inconsistent with superradiant instability for rotating BHs, and hence can be ruled out. This suggests that resolution of the cosmological constant problem using zero point length may not be consistent with GW observations and stability of such zero point length corrected BHs. Following this, we present the relevant waveform model and mismatch analysis, which besides the masses and spins of the binary, also includes the universal quantum gravity parameter $\beta$. 

\section{Waveform model and mismatch analysis}\label{mismatch}

In this section, we first present the waveform model, which includes the effect from the zero point length contributing to the phasing of the GW waveform through the tidal heating. Using this modified GW waveform we try to provide the mismatch analysis between GW waveforms with the quantum gravity parameter $\beta$, and the GW waveforms associated with classical BHs, which may help in assessing the possibility of constraining $\beta$ using present as well as future generations of GW detectors. We restrict ourselves to comparable mass binary system in this section, while specializing to the case of EMRIs in the next one. 

\subsection{The GW waveform with zero point length}

We will work exclusively in the frequency domain for the rest of the analysis, and in the frequency domain, the GW waveform emanated by binary BH inspiral with equal/near-equal mass ratio reads,
\begin{align}
\widetilde{h}(f)=\mathcal{C}\mathcal{A}(f)\exp\left[\phi_{\rm test}+\delta \phi\right]~,
\end{align}
where, $\mathcal{C}$ is an overall normalization constant, dependent on the source location, its orientation and antenna pattern functions. The frequency-dependent amplitude $\mathcal{A}(f)$, appearing in the GW waveform, in the inspiral regime reads: $\mathcal{A}(f)\sim D_{\rm L}^{-1}M_{\rm c}^{5/6}f^{-7/6}$, where, $D_{\rm L}$ is the Luminosity distance and the chirp mass $M_{\rm c}$ of the binary is given by, $M_{\rm c}=(M_{1}M_{2})^{3/5}(M_{1}+M_{2})^{-1/5}$. Finally, the GW phase, has been divided into two parts: $\phi_{\rm test}$ is the phase contribution arising from the point particle limit, and $\delta \phi_{\rm zero}$ is the contribution to the phase due to tidal heating of BHs with zero point length. Among these two terms, the point particle contribution to the GW phase, i.e., $\phi_{\rm test}$, is accurate upto $\mathcal{O}(3.5~\textrm{PN})$, while the tidal heating part, namely $\delta \phi$ has terms involving $2.5$ PN, $3.5$ PN and $4$ PN. Thus the effect due to tidal heating indeed arises at a lower PN order and at that same order it captures the effect of zero point length. Thus the above GW waveform should accurately describe the GW released from the inspiral of aligned spin Kerr BHs with zero point length.  

To explicitly depict the GW phasing due to tidal heating, we present herewith the phasing formula for GWs, which under the stationary phase approximation reads \cite{Tichy:1999pv}, 
\begin{align}
\phi(v)=-2\int^{v}d\mathtt{v}\left(v^{3}-\mathtt{v}^{3}\right)\frac{(dE_{\rm orb}/d\mathtt{v})}{\mathcal{F}_{\infty}+\mathcal{F}_{\rm H}}~,
\end{align}
where, $E_{\rm orb}$ is the orbital energy, $\mathcal{F}_{\infty}$ is the GW flux at infinity and $\mathcal{F}_{\rm H}$ is GW flux through the horizon. Moreover, $\phi(v)$ depicts the total phase accumulated as the two BHs inspiral around each other reaching a relative velocity $v$. Using the quadrupolar formula for GW flux at infinity, given by $\mathcal{F}_{\infty}$, as well as the expression for orbital energy and its derivative, along with the flux through the horizon $\mathcal{F}_{\rm H}$, the phasing due to tidal heating of BHs with zero point length yields \cite{Krishnendu:2024jkj},
\begin{widetext}
\begin{align}\label{phaseQBH}
\delta \phi^{\rm QBH}&=\frac{10}{32\eta}\int^{v}d\mathtt{v}\Bigg[\left(\frac{v^{3}-\mathtt{v}^{3}}{\mathtt{v}^{4}}\right)\frac{\Psi_{5}}{4}+\left(\frac{995}{168}+\frac{952}{168}\eta\right)\left(\frac{v^{3}-\mathtt{v}^{3}}{\mathtt{v}^{2}}\right)\frac{\Psi_{5}}{4}
-\left(\frac{v^{3}-\mathtt{v}^{3}}{\mathtt{v}}\right)\left\{\frac{\Psi_{8}}{2}+\frac{\Psi_{5}}{2}\left(4\pi+F_{\rm SO}\right)\right\}\Bigg]
\nonumber
\\
&=\frac{10}{32\eta}\sum_{i=1}^{2}\Bigg[-\frac{\Psi_{5(i)}^{\rm (BH)}}{12}\left(1+3\ln v -3\ln v_{i(\rm min)}\right)+\left(\frac{995}{168}+\frac{952}{168}\eta\right)\frac{\Psi_{5(i)}^{\rm (BH)}}{8}v_{i(\rm min)}^{2}
\nonumber
\\
&-\left\{\frac{\Psi_{8(i)}^{\rm (BH)}}{6}+\frac{\Psi_{5(i)}^{\rm (BH)}}{6}\left(4\pi+F_{\rm SO}\right)\right\}v_{i(\rm min)}^{3}
-\left(\frac{995}{168}+\frac{952}{168}\eta\right)\frac{3\Psi_{5(i)}^{\rm (BH)}}{8}v^{2}
+\frac{\Psi_{5(i)}^{\rm (BH)}v^{3}}{12~v_{i(\rm min)}^{3}}
+\left(\frac{995}{168}+\frac{952}{168}\eta\right)\frac{\Psi_{5(i)}^{\rm (BH)}v^{3}}{4 v_{i(\rm min)}}
\nonumber
\\
&\qquad +\left\{\frac{\Psi_{8(i)}^{\rm (BH)}}{6}+\frac{\Psi_{5(i)}^{\rm (BH)}}{6}\left(4\pi+F_{\rm SO}\right)\right\}\left(1-3\ln v +3\ln v_{i(\rm min)}\right)v^{3}\Bigg]
\nonumber
\\
&=\frac{10}{32\eta}\sum_{i=1}^{2}\Bigg[\frac{\Psi_{5(i)}^{\rm (BH)}}{12}\bigg(\frac{v^3}{v_{i(\rm min)}^3}- 1 -\ln\frac{v^3}{v_{i(\rm min)}^{3}}\bigg)
+\bigg(\frac{995}{168} + \frac{952}{168} \eta\bigg)\frac{\Psi_{5(i)}^{\rm (BH)}}{8}v_{i(\rm min)}^{2}\bigg(1-3\frac{v^2}{v_{i(\rm min)}^{2}}+2\frac{v^3}{v_{i(\rm min)}^{3}}\bigg)
\nonumber
\\
&\qquad+\bigg(\frac{\Psi_{8(i)}^{\rm (BH)}}{6}+\frac{\Psi_{5(i)}^{\rm (BH)}}{6}\left(4\pi+F_{\rm SO}\right)\bigg)v_{i(\rm min)}^{3}\bigg(\frac{v^3}{v_{i(\rm min)}^{3}}-1-\frac{v^3}{v_{i(\rm min)}^{3}}\ln\frac{v^3}{v_{i(\rm min)}^{3}}\bigg)
 \Bigg]~.
\end{align}
Here, the quantities $\Psi_{5(i)}^{\rm (BH)}$ and $\Psi_{8(i)}^{\rm (BH)}$ has been defined earlier, in \ref{Psi5bh} and \ref{Psi8bh}, respectively, $\eta$ is the dimensionless mass ratio, defined above \ref{flux_hor}, and $F_{\rm SO}$ is the spin orbit coupling appearing at $1.5$ PN, beyond the leading order, in the GW flux at infinity. Note that the GW phasing depends on the existence of zero point length through $v_{\rm min}=(M\omega_{\rm min}/2)^{1/3}$, which is the minimum velocity beyond which the BHs with zero point length behaves as a classical BH, with $\omega_{\rm min}$ being presented in \ref{minfreq}. The two addends under the summation symbol must be understood with $v \ge v_{i(\rm min)}$ and, by construction, vanish for $v \le v_{i(\rm min)}$.

Given the above formula for GW phasing, and as advocated before, the presence of the zero point length is through the minimum velocity $v_{\rm min}$, which depends on the minimum frequency $\omega_{\rm min}$, and in turn on the zero point length parameter $\beta$. For classical BHs, on the other hand, there is a lower velocity, which corresponds to the frequency threshold of the GW detector. For example, Advanced LIGO type detectors have the lower frequency cutoff at $\sim 20$ Hz. Therefore, the dephasing $\Delta \phi$ between classical and zero point length corrected quantum BHs arise from the existence of a minimum frequency for quantum corrected BHs, satisfying $f_{\rm min}>20$ Hz, and is given by, 
\begin{align}
\Delta \phi&\equiv\delta \phi^{\rm CBH}-\delta \phi^{\rm QBH}=\frac{10}{32\eta}\sum_{i=1}^{2}\Bigg[\frac{\Psi_{5(i)}^{\rm (BH)}}{4}
\left(\ln v_{20}-\ln v_{i(\rm min)}\right)
+\left(\frac{995}{168}+\frac{952}{168}\eta\right)\frac{\Psi_{5(i)}^{\rm (BH)}}{8}\left(v_{20}^{2}-v_{i(\rm min)}^{2}\right)
\nonumber
\\
&-\left\{\frac{\Psi_{8(i)}^{\rm (BH)}}{6}+\frac{\Psi_{5(i)}^{\rm (BH)}}{6}\left(4\pi+F_{\rm SO}\right)\right\}\left(v_{20}^{3}-v_{i(\rm min)}^{3}\right)
+\frac{\Psi_{5(i)}^{\rm (BH)}}{12}v^{3}\left(\frac{1}{v_{20}^{3}}-\frac{1}{v_{i(\rm min)}^{3}}\right)
\nonumber
\\
&+\left(\frac{995}{168}+\frac{952}{168}\eta\right)\frac{\Psi_{5(i)}^{\rm (BH)}}{4}v^{3}\left(\frac{1}{v_{20}}-\frac{1}{v_{i(\rm min)}} \right)
+\left\{\frac{\Psi_{8(i)}^{\rm (BH)}}{2}+\frac{\Psi_{5(i)}^{\rm (BH)}}{2}\left(4\pi+F_{\rm SO}\right)\right\}\left(\ln v_{20}-\ln v_{i(\rm min)}\right)v^{3}\Bigg]~.
\end{align}
\end{widetext}
Here, $v_{20}$ is the velocity of the binary system associated with the $20$ Hz cut-off of the Advanced LIGO type detectors. Thus, there is a distinct difference between the GW phase arising from the tidal heating of the classical BHs and the tidal heating of zero point length corrected BHs. This difference in the GW phasing should show up in the mismatch analysis involving GW waveform and we would like to discuss implications of this difference for constraining the parameter $\beta$ from future as well as present generations of GW detectors. 

\begin{figure}[htbp]
\includegraphics[width=0.5\textwidth]{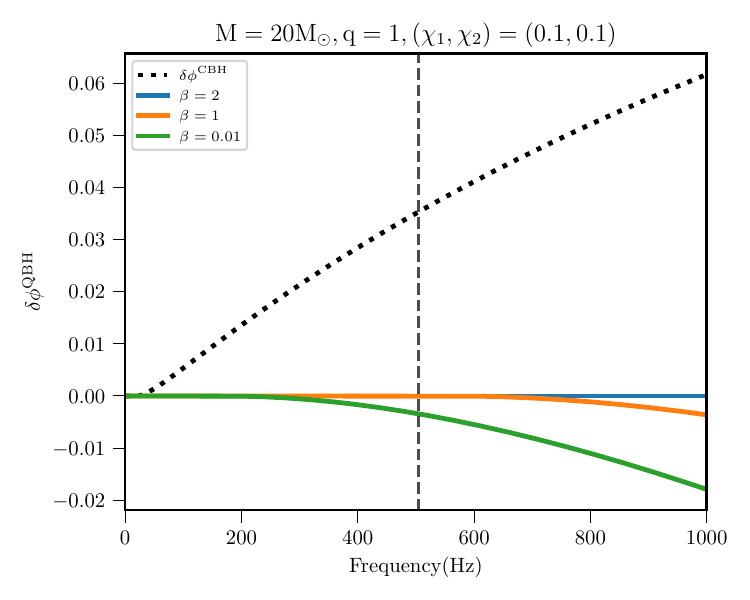}
\caption{The phase $\delta\phi^{\rm QBH}$ due to tidal heating of a BH binary with zero point length has been plotted as a function of the frequency. We considered the BH binary to be symmetric with total mass $M=20M_{\odot}$, mass ratio $q=1$ and spin $\chi_{1}=0.1=\chi_{2}$. The blue, orange and green curves correspond to three different $\beta$ values, namely $\beta=2$, $\beta=1$ and $\beta=0.01$, respectively. For comparison we have also depicted the phase $\delta\phi^{\rm{CBH}}$ of a classical BH binary with same parameters by the black dashed line. The dashed vertical line indicates the corresponding ISCO frequency, which for the present binary BH system is 503.4 Hz.}
\label{fig:qbh_cbh_phase_high_spin}
\end{figure}

\begin{figure}[htbp]
\includegraphics[width=0.5\textwidth]{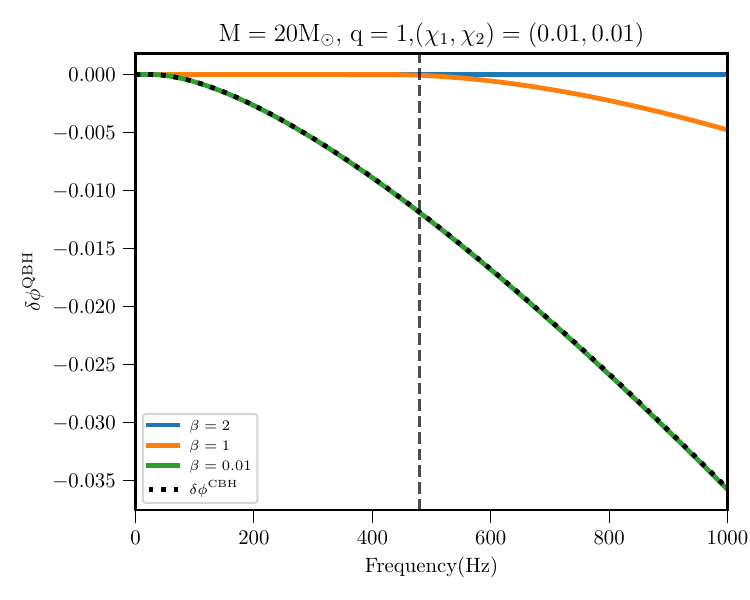}
\caption{The phase of a quantum BH binary, due to tidal heating, is presented against the frequency. The binary BH system is taken to be symmetric, with total mass $M=20 M_{\odot}$, mass ratio $q=1$, and spins $\chi_{1}=\chi_{2}=0.01$. Here also the blue, orange and green curves correspond to three different $\beta$ values, namely $\beta=2$, $\beta=1$ and $\beta=0.01$, respectively. As evident, for $\beta=0.01$, the phasing of the quantum BH follows that of the classical BH. The vertical dashed line depicts the ISCO frequency.}
\label{fig:qbh_cbh_phase_low_spin}
\end{figure}

\begin{figure*}[htbp]
\includegraphics[width=0.8\textwidth]{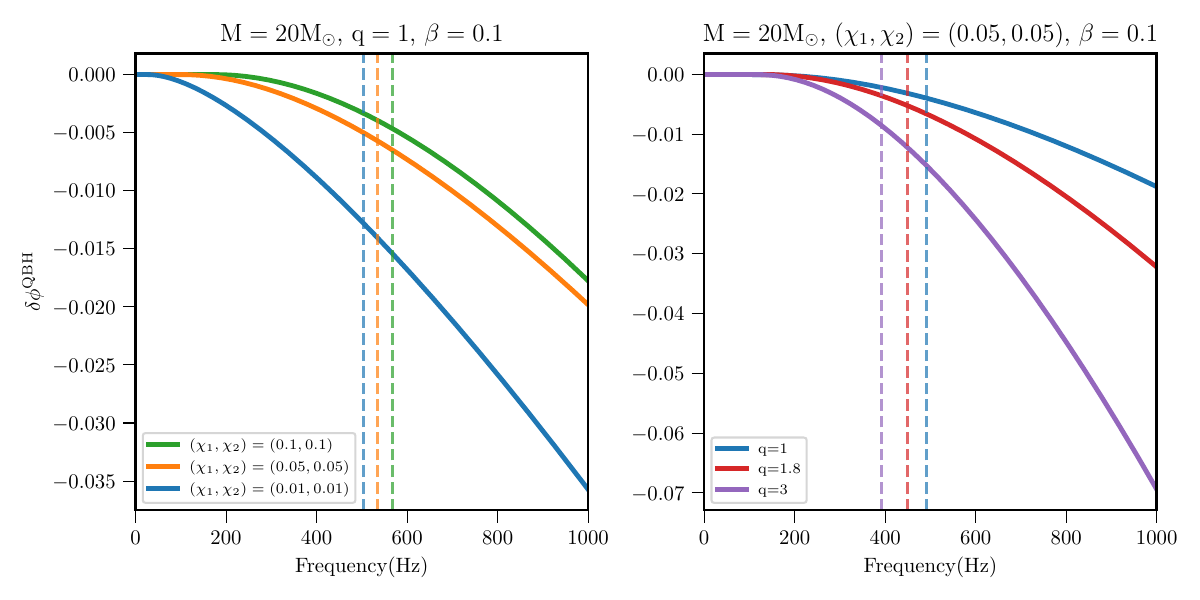}
\caption{Phasing of a quantum BH binary, $\delta\phi^{\rm QBH}$, as a function of the frequency has been presented for a binary with total mass $M=20 M_{\odot}$ and $\beta=0.1$. Left plot: Fixed mass ratio q=1, with dimensionless spins $(0.01, 0.01)$ (blue), $(0.05, 0.05)$ (orange) and $(0.1, 0.1)$ (green). Right: Fixed spins $(0.05, 0.05)$, but varied mass ratio from q=1 (blue), q=1.8 (red) and q=3 (purple). The dashed vertical lines show the ISCO frequencies.} 
\label{fig:deltaPhi_qbh_spins_q}
\end{figure*}

In evaluating $\Delta \phi$, there are several cases to be considered separately depending on the relative values of the velocities, $v$, $v_{1(\rm min)}$, $v_{2(\rm min)}$ and $v_{20}$ appearing in the above expression. As a general rule, from the construction itself it follows that, if $v_{1(\rm min)}$ and $v_{2(\rm min)}< v_{20}$, then $\Delta \phi=0$. The phase difference also vanishes, if $v<v_{20}$, irrespective of the values of $v_{1(\rm min)}$ and $v_{2(\rm min)}$. Therefore, the formula that we have provided above applies when $\{v,v_{1(\rm min)}$, $v_{2(\rm min)}\}\ge v_{20}$. To gain an idea of the numbers involved in the formulas, we can consider the equal mass case, say with, $M_{1}= M_{2}=10 M_\odot$ ($M_\odot$ being the solar mass), and we get $v_{20} \simeq 0.21$; whereas for $f= 500 \, \, {\rm Hz} \approx f_{\rm ISCO}$ ($f_{\rm ISCO}$ is the GW frequency associated to the inner-most stable circular orbits) we get $v_{500} \simeq 0.61$; we have then the relevant range of the velocity $v$ to be $1\le(v/v_{20})\lesssim 3$. In particular, for low spin and small $\beta$ it follows that $\{v_{1(\rm min)}$, $v_{2(\rm min)}\}\le v_{20}$ and hence there is no phase difference between classical and quantum BH. While for intermediate spin we have $\{v_{1(\rm min)}$, $v_{2(\rm min)}\}\ge v_{20}$, and hence there are non-trivial phase differences between classical and quantum BHs. 

Further, notice that the phasing of the GW waveform arising from the inspiral of binary BH systems are influenced by the tidal induced deformations, which we have ignored in the present analysis. This can be justified, since (a) tidal deformation arise in a higher PN order (it starts appearing at the 5 PN order), and (b) classical BHs have vanishing static TLNs, while the zero point length corrected BHs may have non-zero TLNs, but their magnitudes will be small \cite{Nair:2022xfm, Chakraborty:2023zed}. Hence contributions from tidal deformation can safely be ignored.

\subsection{Mismatch between classical and quantum BHs}

Following our analytical expression for the phasing formula for GWs due to tidal heating, we have presented its behaviour with frequency for various choices of the zero-point length parameter $\beta$, along with the phasing for GWs from classical BH binary, in \ref{fig:qbh_cbh_phase_high_spin} and \ref{fig:qbh_cbh_phase_low_spin}, respectively. As \ref{fig:qbh_cbh_phase_high_spin} demonstrates, for higher values of $\beta$, the tidal heating has no contribution to the phasing of the quantum BH in the frequency range of interest. While for smaller values of $\beta$, there are non-trivial contributions to the GW phasing from the tidal heating. Same trend is observed in \ref{fig:qbh_cbh_phase_low_spin} as well. A comparison between \ref{fig:qbh_cbh_phase_high_spin} and \ref{fig:qbh_cbh_phase_low_spin} reveals that the contribution to the phasing due to tidal heating of QBH increases as the rotation decreases. Further, the phasing due to tidal heating of classical BHs is positive for higher spin values, while it becomes negative for smaller spin values. As evident from these plots, the difference between the phasing of Quantum BH and Classical BH is significant at all frequencies of interest for large rotation. While, as evident from \ref{fig:qbh_cbh_phase_low_spin}, for slowly rotating BHs, with small values of $\beta$, the classical and quantum BHs have identical phasing, which is consistent with our findings in \ref{tidalheating}. 

\begin{figure*}[htbp]
\includegraphics[width=0.8\textwidth]{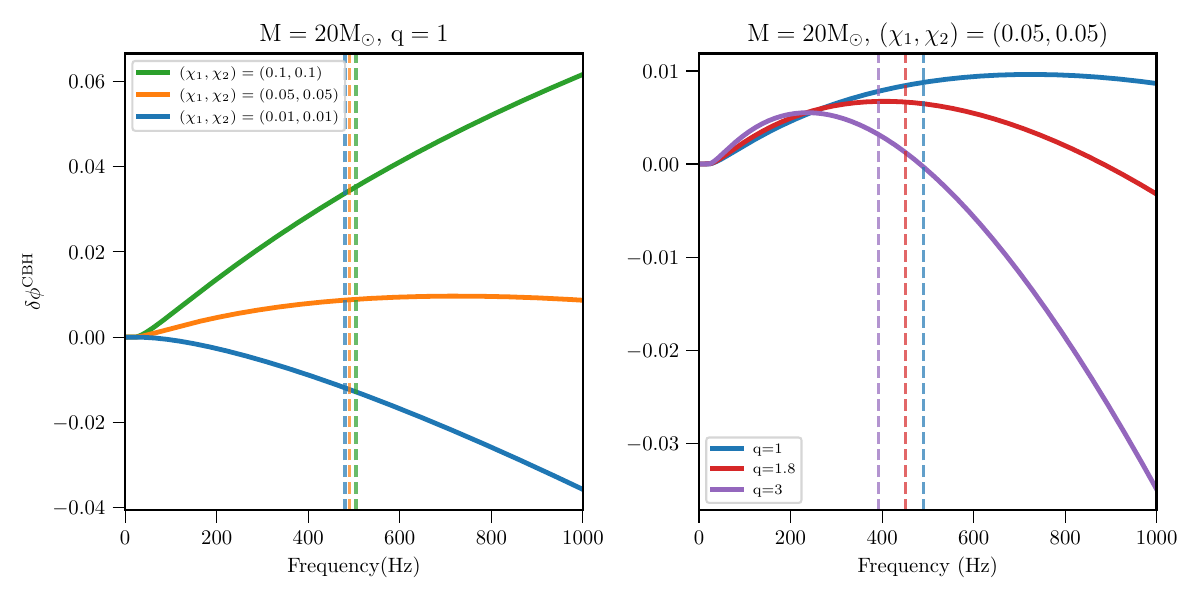}
\caption{Phasing of a classical BH binary with total mass $M=20 M_{\odot}$, $\delta\phi^{\rm CBH}$, has been presented as a function of the frequency. Left plot: Fixed mass ratio q=1, with dimensionless spins $(0.01, 0.01)$ (blue), $(0.05, 0.05)$ (orange) and $(0.1, 0.1)$ (green). Right plot: Fixed spins $(0.05, 0.05)$, but varied mass ratio from q=1 (blue), q=1.8 (red) and q=3 (purple). The dashed vertical lines show the ISCO frequencies. As evident, the phasing due to tidal heating of classical BH binaries can be either positive or negative, depending on the spins and the mass ratio.} 
\label{fig:deltaPhi_cbh_spins_q}
\end{figure*}

\begin{figure*}[htbp]
\includegraphics[width = 0.48\textwidth]{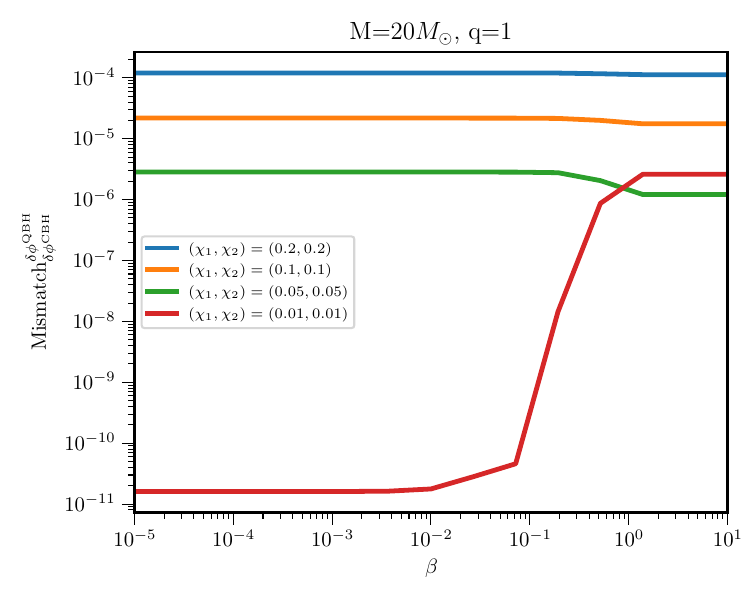}
\includegraphics[width = 0.48\textwidth]{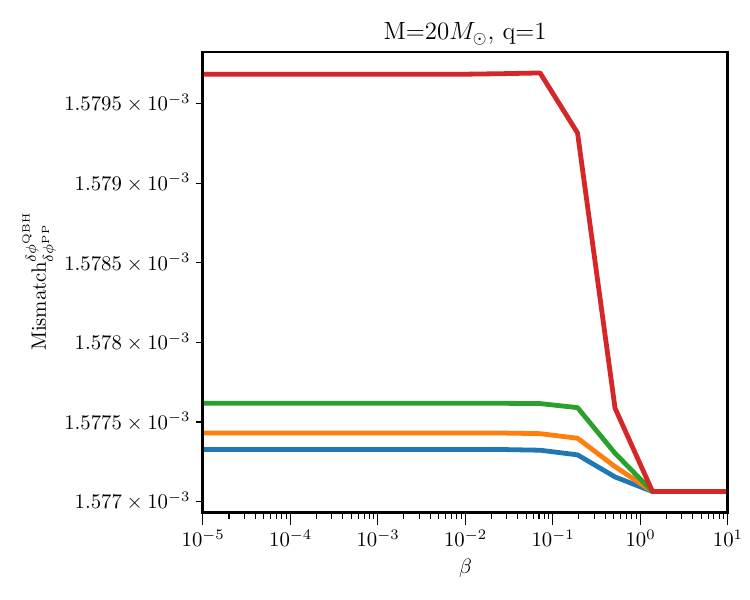}
\caption{Left plot: The mismatch between the waveform model for quantum BH binary, with $\delta\phi^{\rm QBH}$ and classical BH binary, with $\delta\phi^{\rm{CBH}}$ has been presented against the zero point length parameter $\beta$. Right plot: The mismatch between waveform models of quantum BH and a point-particle has been presented against $\beta$. The BH binary is assumed to have total mass $M=20M_{\odot}$ and mass ratio $q=1$ for different dimensions less spins, $\chi_{1}=0.01=\chi_{2}$ (red), $\chi_{1}=0.05=\chi_{2}$ (green), $\chi_{1}=0.1=\chi_{2}$ (orange) and $\chi_{1}=0.2=\chi_{2}$ (blue). As evident, the mismatch between classical and zero point length corrected BH is lowest for small spin and small $\beta$, while the mismatch between point particle and zero point length corrected BH is lowest for larger $\beta$, irrespective of spin.} 
\label{fig:mismatch_cbh_qbh}
\end{figure*}

\begin{figure}[htbp]
\includegraphics[width = 0.5\textwidth]{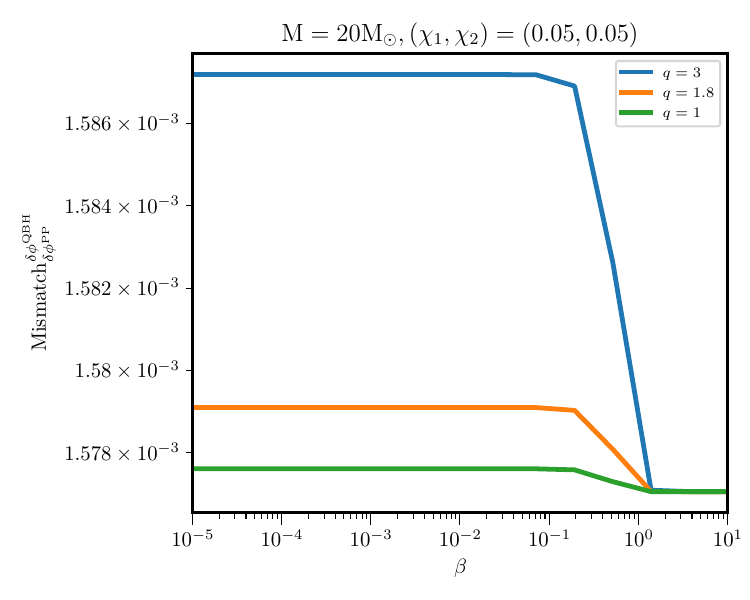}
\caption{The mismatch between waveform model with $\delta\phi^{\rm QBH}$ and a point-particle waveform model, has been presented against zero point length parameter $\beta$, for a fixed dimensionless spin of $\chi_1=0.05= \chi_2$ and total mass $M=20M_{\odot}$. We have depicted the mismatch for three different values of the mass ratio, $q=1$ (green), $q=1.8$ (orange) and $q=3$ (blue). Here the mismatch is lowest for large $\beta$. }
\label{fig:mismatch_pp_qbh_q}
\end{figure}

The variation of the phasing of a quantum BH binary and a classical BH binary has been presented for different spin values, as well as for different mass ratios in \ref{fig:deltaPhi_qbh_spins_q} and \ref{fig:deltaPhi_cbh_spins_q}, respectively. In the case of the quantum BH binary, we have taken $\beta=0.1$, while for both the classical and the quantum BH binary, we have taken the total mass of the binary to be $M=20 M_{\odot}$. From the left hand figure of \ref{fig:deltaPhi_qbh_spins_q}, it is evident that for smaller rotation, the phasing due to tidal heating is much higher than for BH binaries with higher rotation. Similarly, the right hand side figure of \ref{fig:deltaPhi_qbh_spins_q} demonstrates that BH binaries with higher mass asymmetry has higher phasing contribution. This suggests that quantum BH binaries with higher mass asymmetry and lower spin has the highest contribution to the GW phasing due to tidal heating and are the prime candidates to constrain the zero point length parameter $\beta$. For a classical BH binary, on the other hand, for higher rotation parameters the phasing due to tidal heating is positive, while for smaller rotation parameter the phasing is negative, see the left hand figure of \ref{fig:deltaPhi_cbh_spins_q}. Similarly, the right hand figure of \ref{fig:deltaPhi_cbh_spins_q} demonstrates that classical BH binaries with higher mass asymmetry has larger, but negative phasing, while binaries with smaller mass asymmetry has smaller and positive phasing. 


The mismatch between the waveforms constructed using the incorporation of the phasing formula for $\delta\phi^{\rm QBH}$ and $\delta \phi^{\rm CBH}$ in the \texttt{TaylorF2}, which is an aligned spin inspiral waveform model for binary BHs, waveform model has been presented in \ref{fig:mismatch_cbh_qbh}. Besides, we have also provided the mismatch plot between the quantum BH and the point-particle waveform model in both \ref{fig:mismatch_cbh_qbh} and \ref{fig:mismatch_pp_qbh_q}. As evident from the left hand plot of \ref{fig:mismatch_cbh_qbh}, the mismatch between classical and quantum BH binaries is the highest for higher rotation parameter of the individual BHs, but there is no variation with $\beta$. Hence BHs with higher rotation parameter are unable to provide any constraint on the zero point length parameter $\beta$. On the other hand, for smaller rotation parameter, the mismatch is smaller, but its variation with $\beta$ is very large, suggesting that BHs with smaller rotation parameter provides the best case scenario for constraining $\beta$. This feature is also present in the right hand side plot of \ref{fig:mismatch_cbh_qbh}, which shows that the mismatch between the point-particle waveform and the quantum BH binary is the maximum for smaller rotation, which also has the highest variation with $\beta$. Similarly, in \ref{fig:mismatch_pp_qbh_q} we have demonstrated the mismatch between point particle and quantum BH waveform with the zero-point length parameter $\beta$ and the mass ratio $q$. As evident the mismatch is highest for asymmetric BH binary, which also has the highest variation with $\beta$. Thus low spin, asymmetric BH binaries are the best bet to get an estimate for the zero-point length parameter $\beta$. Note that $\beta$ is an universal parameter and does not depend on the individual BH parameters.

Even though it is clear that there exist mismatch between the GW waveforms involving quantum BH and point particles, the amount of mismatch is a small quantity. In particular, as evident from \ref{fig:mismatch_pp_qbh_q}, the mismatch is $\mathcal{O}(10^{-4})$ for the best case scenario, i.e., for slowly rotating BH binaries with asymmetry. From such a small mismatch it is difficult to obtain any significant constraint on $\beta$ through the present parameter inference techniques involving Bayesian analysis. Thus, even though we have performed a GW parameter inference analysis by employing Bayesian techniques, the result was an unconstrained posterior for $\beta$. This suggests that using comparable mass binary BH systems, at least using present generations of GW detectors, it is difficult to provide any reasonable constraint on $\beta$. 
\section{Zero point length and extreme mass ratio inspiral}\label{EMRI}

\begin{figure*}[t]
\includegraphics[width = 0.48\textwidth]{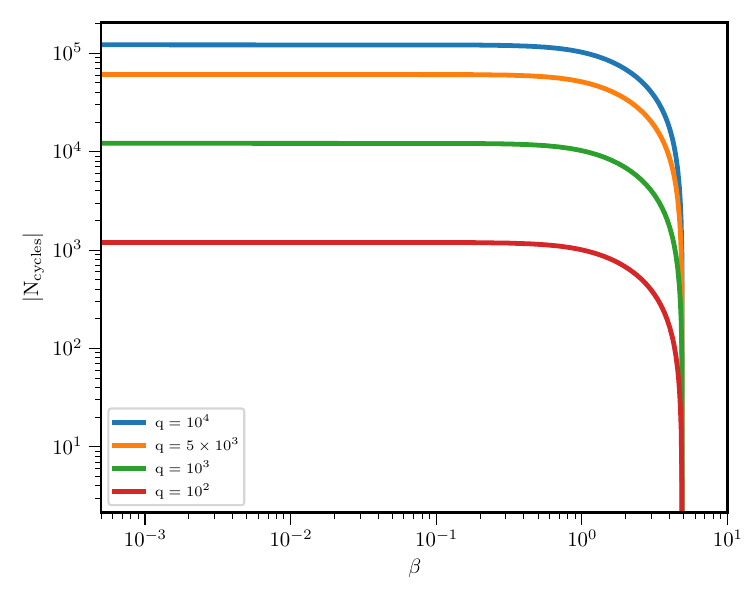}
\includegraphics[width = 0.48\textwidth]{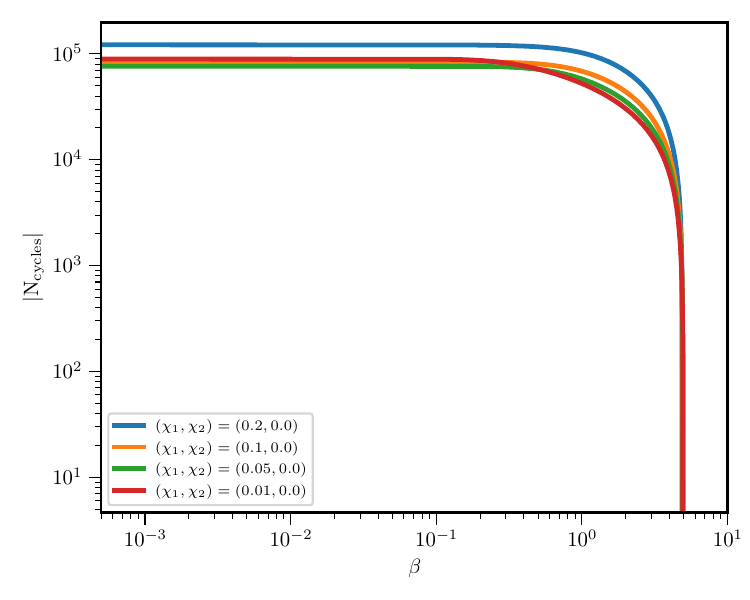}
\caption{The number of cycles associated with the phasing of GW due to tidal heating of zero point length corrected primary BH in an EMRI has been presented against the zero point length parameter $\beta$. We start with the lower cut off frequency of the LISA detector, located at $0.5$ mHz and extend it upto the corresponding ISCO frequency of the primary Kerr BH. The plot on the left hand side depicts the variation of the number of cycles with the mass ratio $q$ for an EMRI with total mass of $10^{6}M_{\odot}$ and primary spin $\chi=0.2$. The right hand side plot depicts variation of the number of cycles for different spins of the primary BH in an EMRI with total mass of $10^{6}M_{\odot}$ and mass ratio $q=10^{4}$.}
\label{fig:f_min_alpha_spins_EMRI2}
\end{figure*}

As we have depicted before, using comparable mass binary BH systems, it is really difficult to constrain the dimensionless ratio between the zero point length and the Planck length. Thus we consider in this section the case of EMRI, both for completeness, as well as to show how the phase difference due to tidal heating can be used to constrain $\beta$. We will assume that the secondary object has a mass much smaller than the primary, i.e., $M_{2}\ll M_{1}$, typical of any EMRI system and hence the total mass of the system is predominantly determined by $M_{1}$ alone ($M\approx M_{1}$). Therefore, in this limit, from \ref{Psi5bh} and \ref{Psi8bh}, it follows that both $\Psi_{5}$ and $\Psi_{8}$ becomes negligibly small for the secondary object, while for the primary these are given by, 
\begin{align}
\Psi^{\rm BH}_{5\textrm{(primary)}}&=\chi(1+3\chi^{2})\left(\hat{\mathbf{L}}_{\rm orb}\cdot \hat{\mathbf{J}}\right)~,
\\
\Psi^{\rm BH}_{8\textrm{(primary)}}&=(1+3\chi^{2})\left(1+\sqrt{1-\chi^{2}}\right)~,
\end{align}
where, all the quantities refer to that of the primary BH. The relative velocity will effectively become the velocity of the secondary, as in EMRI, the primary can be taken to be at rest. Just as in the context of LIGO observatories, the minimum observable frequency corresponds to $20$ Hz, for LISA, the GW observatory that will detect GWs from EMRIs, the minimum frequency will correspond to $0.5$ mHz. The associated velocity will be denoted by $v_{0.5}$. Similarly, the minimum relative velocity between the primary and secondary for which the primary object will start absorbing GW is given by, 
\begin{align}
v_{\rm min}=\frac{1}{2}\,\left(\frac{\beta^{2}\sqrt{1-\chi^{2}} + 4\chi}{1+\sqrt{1-\chi^{2}}}\right)^{1/3}~.
\end{align}
Therefore, the phase difference between the point-particle and the zero-point length corrected BH is given by precisely the tidal heating term, which can be obtained for EMRI from \ref{phaseQBH} by simply setting the $\Psi_{5(2)}$ and $\Psi_{8(2)}$ terms to zero, along with expressing $\eta\approx q^{-1}$, where $q=(M_{\rm primary}/M_{\rm secondary})\gg 1$, is the mass ratio. This is because, the point particle waverform does not involve tidal heating and hence any contribution coming from the tidal heating must be due to the zero-point length corrected BH. Therefore, the phase difference between the classical and the zero point length corrected BH, is given by,
\begin{widetext}
\begin{align}\label{emriphase}
\Delta \phi^{\rm EMRI}&=\frac{10 q}{32}\Bigg[\frac{\Psi_{5}^{\rm (BH)}}{4}
\left(\ln v_{0.5}-\ln v_{\rm min}\right)
+\left(\frac{995}{168}\right)\frac{\Psi_{5}^{\rm (BH)}}{8}\left(v_{0.5}^{2}-v_{\rm min}^{2}\right)
\nonumber
\\
&-\left\{\frac{\Psi_{8}^{\rm (BH)}}{6}+\frac{\Psi_{5}^{\rm (BH)}}{6}\left(4\pi+F_{\rm SO}\right)\right\}\left(v_{0.5}^{3}-v_{\rm min}^{3}\right)
+\frac{\Psi_{5}^{\rm (BH)}}{12}v^{3}\left(\frac{1}{v_{0.5}^{3}}-\frac{1}{v_{\rm min}^{3}}\right)
\nonumber
\\
&+\left(\frac{995}{168}\right)\frac{\Psi_{5}^{\rm (BH)}}{4}v^{3}\left(\frac{1}{v_{0.5}}-\frac{1}{v_{\rm min}} \right)
+\left\{\frac{\Psi_{8}^{\rm (BH)}}{2}+\frac{\Psi_{5}^{\rm (BH)}}{2}\left(4\pi+F_{\rm SO}\right)\right\}\left(\ln v_{0.5}-\ln v_{\rm min}\right)v^{3}\Bigg]~.
\end{align}
\end{widetext}
The above expression is valid for $v_{\rm min}>v_{0.5}$, while for $v_{\rm min}\leq v_{0.5}$, the above phasing formula vanishes identically. In other words, if the minimum frequency $f_{\rm min}$ of the supermassive quantum BH acting as the primary source of an EMRI, is smaller than the cut off frequency of LISA (0.5 mHz), then the quantum BH will behave as a classical BH for intents and purposes and hence the phase difference between classical and quantum BH will vanish. While for $f_{\rm min}>0.5$ mHz, it follows that there are phase differences between classical and quantum BHs.

In the context of EMRI, the best way to depict the phase difference is through the computation of the number of cycles, i.e., the number of complete rotation of the secondary around the primary due to tidal heating. Mathematically, this is given by, 
\begin{align}
N_{\rm cycle}(f)=\int^{f_{\rm ISCO}}_{f}dF\frac{F}{\dot{F}}~,
\end{align}
where, $F$ is a dummy variable of the integration and depicts GW frequency, $f_{\rm ISCO}$ is the GW frequency emitted from the EMRI, as the secondary object is at the ISCO of the primary. The time derivative of the frequency, $\dot{F}$, is determined by using the result that $\dot{E}_{\rm orbit}=\dot{E}_{\rm GW}+\dot{E}_{\rm H}$ along with the result that $E_{\rm orbit}=(GM_{1}M_{2}/r^{2})$, and $r=M^{1/3}(\pi f)^{-2/3}$, such that from $\dot{E}_{\rm otbit}$, we can immediately obtain $\dot{r}$ and hence $\dot{f}$, which will be used in the number of cycles computation. Here, $\dot{E}_{\rm GW}$ is the energy loss due to GW emission and $\dot{E}_{\rm H}$ is the energy loss to the BH horizon, i.e., tidal heating. The above integral can also be written as integral of the frequency $f$ over time, and since $\pi f=(d\phi/dt)$, it follows that number of cycles is directly proportional to the phasing $\phi$ of GW. 

We have presented the variation in the number of cycles with $\beta$, due to the tidal heating term associated with quantum BHs for different spins of the individual BHs and different mass ratios in \ref{fig:f_min_alpha_spins_EMRI2}. The left hand plot in \ref{fig:f_min_alpha_spins_EMRI2} shows the variation of the number of cycles with $\beta$, due to different mass ratios between primary and secondary. As evident, irrespective of the mass ratio, for smaller $\beta$, we have large modifications to the number of cycles. This is because, the quantum BH behaves as a classical BH, for frequencies about the cut off frequency of LISA, and hence produces a significant contribution due to tidal heating. Further, as \ref{emriphase} demonstrates, the phasing due to tidal heating is directly proportional to the mass ratio $q$. Therefore, as the mass ratio becomes larger, the number of cycles also increases dramatically. Also for $0.1\lesssim\beta\lesssim2$, the number of cycles changes drastically, and for $\beta\gtrsim 2$, the number of cycles vanishes very fast. This again suggest that the range $0.1\lesssim \beta \lesssim 2$ is the golden window to constrain $\beta$, of course, the existence of non-zero $\beta$ can be inferred for any $\beta\lesssim 4$, since the number of cycles due to the existence of zero point length satisfies $N_{\rm cycle}\gtrsim 1$ for any $\beta\lesssim 4$. A similar conclusion can be drawn from the right hand side plot of \ref{fig:f_min_alpha_spins_EMRI2} as well. For smaller $\beta$, the number of cycles due to tidal heating is large, as the quantum BH behaves essentially as classical, while for $\beta \gtrsim 4$, the number of cycles drops to zero, as the quantum BH becomes perfectly reflecting, leading to vanishing contribution to the tidal heating. It seems that the spins do not have a significant effect on the number of cycles computation. 
\section{Concluding remarks}\label{conc}

The main results of this work can be summarized as follows:
both for equal mass stellar mass BH binaries and for EMRIs, the existence of zero point length modifies the GW waveform, which at the leading PN order appears through the tidal heating term in the inspiral. 
It follows that, the ratio of the zero point length $\ell_{0}$ and the Planck length $\ell_{\rm p}$ acts as the free parameter of these quantum corrected theories, which we denote as $\beta\equiv (\ell_{0}/\ell_{\rm p})$.
The existence of zero point length prohibits the absorption of GWs of arbitrary frequency by the BH horizon. 
This is because, the increase in the area of the BH horizon can be thought of as increasing the affine distance between two points along the null generators on the BH horizon, and the affine distance is bounded from below by the zero point length. 
As a consequence, the change in the area of a BH horizon is bounded from below by $4\pi \ell_{0}^{2}$, which leads to a minimum frequency $\omega_{\rm min}$ for GWs that the BH can absorb. 
The existence of a minimum frequency, also requires quantization of angular momentum and assuming that the dominant GW mode corresponds to $\ell=2=m$. 
The minimum frequency $\omega_{\rm min}$ depending on the angular velocity of the BH horizon, its surface gravity, and of course $\beta$, which is the ratio of the zero point length and the Planck length. 

The consequence of the existence of a minimum frequency is manyfold. First of all, if the minimum frequency is larger than the ISCO frequency ($f_{\rm min}\geq f_{\rm ISCO}$), then it follows that, throughout the inspiral, the BH will be perfectly reflective, and hence the most dominant effect in the GW waveform will be the absence of tidal heating.   
On the other hand, if $f_{\rm min}\lesssim f_{\rm cut~off}$, where, $f_{\rm cut~off}$ is $20$Hz for advanced LIGO like detetors and $0.5$ mHz for LISA, then it follows that the zero point length corrected BHs will behave as classical BH for all intents and purposes. 

Following these broad results, we arrived at the following landscape in the $(\beta,\chi)$ plane, where $\chi\equiv(J/M^{2})$ is the dimensionless angular momentum: 
(a) for small values of spin ($\chi \lesssim 0.02$)
and limit length ($\beta \lesssim 0.1$) the BHs (both the solar mass ones, and the supermassive ones) are perfectly absorbing, i.e., they behave as classical black holes.
(b) At the other extreme, when we have high spin ($\chi \gtrsim 0.25$) for any $\beta$, or high $\beta$ ($\beta \gtrsim 2$) for any spin, we have perfect reflectivity which shows that we are maximally away from the classical scenario.
(c) In the intermediate cases, i.e., small or intermediate spin with intermediate $\beta$ ($0.1\lesssim\beta \lesssim 2$), or for intermediate spin with small $\beta$, we have varying levels of reflectivity depending on the values of spin and $\beta$ (cf. \ref{tab1}).

An intriguing consequence of what we have just described is that, if the coalescence of solar mass BHs, or EMRIs, with very high spin parameters for the individual BHs are detected, and the reconstructed (posterior) reflectivity of these BHs from the inspiral part of the GW waveform turns out to be smaller than unity, then this would be quite a strong indication of a tension with gravity being quantum.
This is because, for high spin, the quantization of the angular momentum of a BH demands the BH to be perfectly reflective.
This shows that highly spinning BHs are important candidates for assessing the quantum nature of gravity, as by analysing the GW signals originating from the inspiral phases of their coalescences (either by the LVK detectors, or by the future space based LISA mission), one can extract the reflectivity of these highly spinning BHs, and depending on the value of the reflectivity, we can confirm or reject the quantum nature for gravity.

On the other hand, for $0.1\lesssim \beta \lesssim 2$, the slowly rotating BHs are partially reflective, implying that they have differences from the classical BHs as well as from a point particle description. Thus from the mismatch between the GW waveform of these partially reflecting BHs with the classical ones, it is possible to constraint $\beta$. For the LVK detectors, it turns out that such a constraint on $\beta$ is hard to obtain, as the signal to noise ratio is very poor, however, from future space based mission, e.g., LISA, the prospect of constraining $\beta$ is significant, as we have depicted through a number of cycles comparson. In particular, $\beta\lesssim 4$ can be well constrained using EMRIs detected in LISA.  
%


There are several future directions of exploration. We have only considered the effect of zero point length on the tidal heating, while it would be interesting to study the effect of zero point length on tidal deformations as well. In particular, it is expected that the dynamical tidal Love numbers will be zero for $f>f_{\rm min}$, but will show a logarithmic behaviour for lower frequencies. Moreover, implications of having such BHs in a DM environment can also provide us important clues about the quantum nature of these BHs, as the GW phasing depends on the accretion, which in turn depends on the reflectivity of the compact objects. Finally, it would be really interesting to devise a more sensitive parameter estimation technique, so that the small mismatch between BHs with zero point length against the classical BHs can be utilized in obtaining a bound on the zero point length. With future GW detectors as well the situation is supposed to improve significantly \cite{Ajith:2024inj, Borhanian:2022czq, ET:2019dnz}.

\section*{Acknowledgments}

The research of SC is supported by MATRICS (MTR/2023/000049) and Core Research Grants (CRG/2023/000934) from SERB, ANRF, Government of India. SC also thanks the local hospitality at ICTS and IUCAA through the associateship program, where a part of this work was done. In particular, a part of this research was performed at the ICTS school ``Beyond the Black Hole Paradigm".
NVK is supported by STFC grant ST/Y00423X/1. NVK also thanks the ICTS, Banglore, India for the hospitality during the initial stages of the project. NVK acknowledges the support from SERB for the National postdoctoral fellowship  (Reg. No. PDF/2022/000379) during the time in ICTS. 
The research of Ale.P. is partially supported by INFN grant FLAG. Ale.P. also would like to acknowledge the contribution of the COST action CA23130. This document has LIGO preprint number {\tt LIGO-P2500355}.

\appendix
\labelformat{section}{Appendix #1} 
\labelformat{subsection}{Appendix #1}
\section{Heuristic derivation of the limiting area increment}\label{app:Heuristic}

In this appendix, we provide a heuristic derivation of the minimum area increment obtained from the qmetric approach, from basic principles. 
For this purpose, we start by considering the Rindler frame $\cal R$, with Rindler horizon, which can locally represent the horizon of a Schwarzschild or Kerr BH (see, e.g., \cite{Paddy_book}). 
Let $\xi^{\mu}$ be the Killing vector corresponding to translations in the Rindler time, namely to the Killing time. 
Then the combination $J_{\alpha}=T_{\alpha \beta}\xi^{\beta}$ is the conserved energy-momentum current as perceived by the accelerated observer. 
Integrating such current over a local patch $\Sigma$ of the future Rindler horizon with spatial cross-section $S$, we obtain the total energy crossing this surface $\Sigma$ as,
\begin{eqnarray}
\Delta Q=\int_{\Sigma} J_{\alpha} d\Sigma^{\alpha}=\int_{S} dydz\int_0^\infty dx T_{\mu \nu}\xi^{\nu}l^{\mu}~.
\end{eqnarray}
Here $x=0$ depicts the horizon in local Rindler coordinates, with $l^{\mu}=(1, 1, 0, 0)$ is the affinely parametrized generators of the horizon, with the $x$ coordinate being taken as the affine parameter. Moreover, the coordinates $(y,z)$ describes the cross-section of the Rindler horizon. 
%
On the horizon, the Killing vector $\xi^{\mu}$ can be related to the null generator $l^{\mu}$ through the following relation: $\xi^{\mu}=\kappa xl^{\mu}$, where $\kappa$ is the surface gravity, and the above relation follows from the condition: $\xi^{\mu}\nabla_{\mu}\xi^{\alpha}=\kappa \xi^{\alpha}$. Thus, the above equation for heat flux through a patch of the Rindler horizon, can be expressed in the following form:
\begin{equation}
\Delta Q=\kappa \int_S dydz \int_0^\infty dx \, x \, T_{\mu \nu}l^{\mu} l^{\nu}~.
\end{equation}
On the other hand, from the Raychaudhuri equation, as applied to the horizon in affine parametrization, and using Einstein's equations, one obtains
\begin{eqnarray}\label{AxT}
\Delta A  = 8\pi \int_S dydz \int_0^\infty dx \, x \, T_{\mu \nu}l^{\mu} l^{\nu}~,
\end{eqnarray}
yielding the following relation between the energy absorbed by the null surface and the area change \cite{Jacobson:1995ab}:
\begin{equation} \label{eq}
\Delta A = \frac{8\pi}{\kappa} \Delta Q~.
\end{equation}


Consider now the situation, in which some matter, say a particle, is dropped at $t=0$ by the observer who has acceleration $\kappa$, when instantaneously at rest in the inertial frame, i.e.,  at $x=(1/\kappa)$. 
%
%
%
In order to include quantum effects into the above calculation, we advocate the following: As a basic tenet of quantum physics we assume that a quantum of energy is either entirely engulfed by the horizon or it is not engulfed at all; that is we forbid that, e.g. a photon, can be in a situation in which it is only partially engulfed.
This amounts to say that, if $\Delta x$ is the uncertainty in the position of the particle, we must have $\Delta x<(1/\kappa)$ (cf. \cite{API}).
This yields the following inequality:
\begin{eqnarray}
\frac{\hbar}{2} \le\Delta p \Delta x \le \frac{\Delta p}{\kappa}
=\frac{\Delta Q}{\kappa}~,  
\end{eqnarray}
where, we have heuristically assumed that the spread in momentum 
to roughly account for the total energy absorbed by the horizon. 
Using this in \ref{eq} we obtain,
\begin{equation}
\Delta A \gtrsim 4\pi  \ell_{\rm p}^2~,
\end{equation}
where, we have restored all the fundamental constants in the last step. 
Thus we recover the qmetric result, provided we identify $\ell_0 = \ell_{\rm p}$, or $\beta = 1$.
This result follows from 
Einstein's equations
and a basic tenet of quantum physics, namely the uncertainty principle. It is worth emphasizing that the very same ingredients bring forward the existence of a zero-point length.

From conservation of energy, this limiting area increment leads to a limiting increment $\Delta M$ to the mass, which gives in turn a minimum frequency which can be absorbed for modes impinging on the hole as described in the main text.
As a case of interest, we present here the minimum frequency with $\beta = 1$, using the result presented in \ref{minfreq}. In the case of Schwarzschild BH, the associated minimum frequency yields,
\begin{eqnarray}
\omega^{\rm Sch}_{\rm min}=\frac{\kappa}{2}=\frac{1}{8M}~,
\end{eqnarray}
and more generally for Kerr, we obtain,
\begin{eqnarray}
\omega_{\rm min}^{\rm Kerr}=\frac{1}{4M} \ \frac{\sqrt{1-\chi^2} + 4 \, \chi}{1 + \sqrt{1-\chi^2}}~,
\end{eqnarray}
where $\chi\equiv (J/M^{2}$) with $M$, and $J$ being the mass and the angular momentum of the BH.  

\bibliography{References}

\begin{thebibliography}{90}%
\makeatletter
\providecommand \@ifxundefined [1]{%
 \@ifx{#1\undefined}
}%
\providecommand \@ifnum [1]{%
 \ifnum #1\expandafter \@firstoftwo
 \else \expandafter \@secondoftwo
 \fi
}%
\providecommand \@ifx [1]{%
 \ifx #1\expandafter \@firstoftwo
 \else \expandafter \@secondoftwo
 \fi
}%
\providecommand \natexlab [1]{#1}%
\providecommand \enquote  [1]{``#1''}%
\providecommand \bibnamefont  [1]{#1}%
\providecommand \bibfnamefont [1]{#1}%
\providecommand \citenamefont [1]{#1}%
\providecommand \href@noop [0]{\@secondoftwo}%
\providecommand \href [0]{\begingroup \@sanitize@url \@href}%
\providecommand \@href[1]{\@@startlink{#1}\@@href}%
\providecommand \@@href[1]{\endgroup#1\@@endlink}%
\providecommand \@sanitize@url [0]{\catcode `\\12\catcode `\$12\catcode
  `\&12\catcode `\#12\catcode `\^12\catcode `\_12\catcode `\%12\relax}%
\providecommand \@@startlink[1]{}%
\providecommand \@@endlink[0]{}%
\providecommand \url  [0]{\begingroup\@sanitize@url \@url }%
\providecommand \@url [1]{\endgroup\@href {#1}{\urlprefix }}%
\providecommand \urlprefix  [0]{URL }%
\providecommand \Eprint [0]{\href }%
\providecommand \doibase [0]{http://dx.doi.org/}%
\providecommand \selectlanguage [0]{\@gobble}%
\providecommand \bibinfo  [0]{\@secondoftwo}%
\providecommand \bibfield  [0]{\@secondoftwo}%
\providecommand \translation [1]{[#1]}%
\providecommand \BibitemOpen [0]{}%
\providecommand \bibitemStop [0]{}%
\providecommand \bibitemNoStop [0]{.\EOS\space}%
\providecommand \EOS [0]{\spacefactor3000\relax}%
\providecommand \BibitemShut  [1]{\csname bibitem#1\endcsname}%
\let\auto@bib@innerbib\@empty
\bibitem [{\citenamefont {Abbott}\ \emph {et~al.}(2016)\citenamefont {Abbott}
  \emph {et~al.}}]{LIGOScientific:2016aoc}%
  \BibitemOpen
  \bibfield  {author} {\bibinfo {author} {\bibfnamefont {B.~P.}\ \bibnamefont
  {Abbott}} \emph {et~al.} (\bibinfo {collaboration} {LIGO Scientific,
  Virgo}),\ }\href {\doibase 10.1103/PhysRevLett.116.061102} {\bibfield
  {journal} {\bibinfo  {journal} {Phys. Rev. Lett.}\ }\textbf {\bibinfo
  {volume} {116}},\ \bibinfo {pages} {061102} (\bibinfo {year} {2016})},\
  \Eprint {http://arxiv.org/abs/1602.03837} {arXiv:1602.03837 [gr-qc]}
  \BibitemShut {NoStop}%
\bibitem [{\citenamefont {Abbott}\ \emph {et~al.}(2019)\citenamefont {Abbott}
  \emph {et~al.}}]{LIGOScientific:2019fpa}%
  \BibitemOpen
  \bibfield  {author} {\bibinfo {author} {\bibfnamefont {B.~P.}\ \bibnamefont
  {Abbott}} \emph {et~al.} (\bibinfo {collaboration} {LIGO Scientific,
  Virgo}),\ }\href {\doibase 10.1103/PhysRevD.100.104036} {\bibfield  {journal}
  {\bibinfo  {journal} {Phys. Rev. D}\ }\textbf {\bibinfo {volume} {100}},\
  \bibinfo {pages} {104036} (\bibinfo {year} {2019})},\ \Eprint
  {http://arxiv.org/abs/1903.04467} {arXiv:1903.04467 [gr-qc]} \BibitemShut
  {NoStop}%
\bibitem [{\citenamefont {Abbott}\ \emph
  {et~al.}(2021{\natexlab{a}})\citenamefont {Abbott} \emph
  {et~al.}}]{LIGOScientific:2020tif}%
  \BibitemOpen
  \bibfield  {author} {\bibinfo {author} {\bibfnamefont {R.}~\bibnamefont
  {Abbott}} \emph {et~al.} (\bibinfo {collaboration} {LIGO Scientific,
  Virgo}),\ }\href {\doibase 10.1103/PhysRevD.103.122002} {\bibfield  {journal}
  {\bibinfo  {journal} {Phys. Rev. D}\ }\textbf {\bibinfo {volume} {103}},\
  \bibinfo {pages} {122002} (\bibinfo {year} {2021}{\natexlab{a}})},\ \Eprint
  {http://arxiv.org/abs/2010.14529} {arXiv:2010.14529 [gr-qc]} \BibitemShut
  {NoStop}%
\bibitem [{\citenamefont {Abbott}\ \emph
  {et~al.}(2021{\natexlab{b}})\citenamefont {Abbott} \emph
  {et~al.}}]{LIGOScientific:2021sio}%
  \BibitemOpen
  \bibfield  {author} {\bibinfo {author} {\bibfnamefont {R.}~\bibnamefont
  {Abbott}} \emph {et~al.} (\bibinfo {collaboration} {LIGO Scientific, VIRGO,
  KAGRA}),\ }\href@noop {} {\  (\bibinfo {year} {2021}{\natexlab{b}})},\
  \Eprint {http://arxiv.org/abs/2112.06861} {arXiv:2112.06861 [gr-qc]}
  \BibitemShut {NoStop}%
\bibitem [{\citenamefont {Abbott}\ \emph {et~al.}(2022)\citenamefont {Abbott}
  \emph {et~al.}}]{LVK:2022}%
  \BibitemOpen
  \bibfield  {author} {\bibinfo {author} {\bibfnamefont {R.}~\bibnamefont
  {Abbott}} \emph {et~al.} (\bibinfo {collaboration} {LIGO Scientific, VIRGO,
  KAGRA}),\ }\href@noop {} {\bibfield  {journal} {\bibinfo  {journal} {Astron.
  Astrophys.}\ }\textbf {\bibinfo {volume} {659}},\ \bibinfo {pages} {A84}
  (\bibinfo {year} {2022})},\ \Eprint {http://arxiv.org/abs/2105.15120}
  {arXiv:2105.15120 [astro-ph.HE]} \BibitemShut {NoStop}%
\bibitem [{\citenamefont {Nitz}\ \emph {et~al.}(2023)\citenamefont {Nitz},
  \citenamefont {Kumar}, \citenamefont {Wang}, \citenamefont {Kastha},
  \citenamefont {Wu}, \citenamefont {Sch\"afer}, \citenamefont {Dhurkunde},\
  and\ \citenamefont {Capano}}]{Nitz:2021zwj}%
  \BibitemOpen
  \bibfield  {author} {\bibinfo {author} {\bibfnamefont {A.~H.}\ \bibnamefont
  {Nitz}}, \bibinfo {author} {\bibfnamefont {S.}~\bibnamefont {Kumar}},
  \bibinfo {author} {\bibfnamefont {Y.-F.}\ \bibnamefont {Wang}}, \bibinfo
  {author} {\bibfnamefont {S.}~\bibnamefont {Kastha}}, \bibinfo {author}
  {\bibfnamefont {S.}~\bibnamefont {Wu}}, \bibinfo {author} {\bibfnamefont
  {M.}~\bibnamefont {Sch\"afer}}, \bibinfo {author} {\bibfnamefont
  {R.}~\bibnamefont {Dhurkunde}}, \ and\ \bibinfo {author} {\bibfnamefont
  {C.~D.}\ \bibnamefont {Capano}},\ }\href {\doibase 10.3847/1538-4357/aca591}
  {\bibfield  {journal} {\bibinfo  {journal} {Astrophys. J.}\ }\textbf
  {\bibinfo {volume} {946}},\ \bibinfo {pages} {59} (\bibinfo {year} {2023})},\
  \Eprint {http://arxiv.org/abs/2112.06878} {arXiv:2112.06878 [astro-ph.HE]}
  \BibitemShut {NoStop}%
\bibitem [{\citenamefont {Abbott}\ \emph {et~al.}(2024)\citenamefont {Abbott}
  \emph {et~al.}}]{LIGOScientific:2021usb}%
  \BibitemOpen
  \bibfield  {author} {\bibinfo {author} {\bibfnamefont {R.}~\bibnamefont
  {Abbott}} \emph {et~al.} (\bibinfo {collaboration} {LIGO Scientific,
  VIRGO}),\ }\href {\doibase 10.1103/PhysRevD.109.022001} {\bibfield  {journal}
  {\bibinfo  {journal} {Phys. Rev. D}\ }\textbf {\bibinfo {volume} {109}},\
  \bibinfo {pages} {022001} (\bibinfo {year} {2024})},\ \Eprint
  {http://arxiv.org/abs/2108.01045} {arXiv:2108.01045 [gr-qc]} \BibitemShut
  {NoStop}%
\bibitem [{\citenamefont {Abbott}\ \emph
  {et~al.}(2021{\natexlab{c}})\citenamefont {Abbott} \emph
  {et~al.}}]{LIGOScientific:2020ibl}%
  \BibitemOpen
  \bibfield  {author} {\bibinfo {author} {\bibfnamefont {R.}~\bibnamefont
  {Abbott}} \emph {et~al.} (\bibinfo {collaboration} {LIGO Scientific,
  Virgo}),\ }\href {\doibase 10.1103/PhysRevX.11.021053} {\bibfield  {journal}
  {\bibinfo  {journal} {Phys. Rev. X}\ }\textbf {\bibinfo {volume} {11}},\
  \bibinfo {pages} {021053} (\bibinfo {year} {2021}{\natexlab{c}})},\ \Eprint
  {http://arxiv.org/abs/2010.14527} {arXiv:2010.14527 [gr-qc]} \BibitemShut
  {NoStop}%
\bibitem [{\citenamefont {Olsen}\ \emph {et~al.}(2022)\citenamefont {Olsen},
  \citenamefont {Venumadhav}, \citenamefont {Mushkin}, \citenamefont {Roulet},
  \citenamefont {Zackay},\ and\ \citenamefont {Zaldarriaga}}]{Olsen:2022pin}%
  \BibitemOpen
  \bibfield  {author} {\bibinfo {author} {\bibfnamefont {S.}~\bibnamefont
  {Olsen}}, \bibinfo {author} {\bibfnamefont {T.}~\bibnamefont {Venumadhav}},
  \bibinfo {author} {\bibfnamefont {J.}~\bibnamefont {Mushkin}}, \bibinfo
  {author} {\bibfnamefont {J.}~\bibnamefont {Roulet}}, \bibinfo {author}
  {\bibfnamefont {B.}~\bibnamefont {Zackay}}, \ and\ \bibinfo {author}
  {\bibfnamefont {M.}~\bibnamefont {Zaldarriaga}},\ }\href {\doibase
  10.1103/PhysRevD.106.043009} {\bibfield  {journal} {\bibinfo  {journal}
  {Phys. Rev. D}\ }\textbf {\bibinfo {volume} {106}},\ \bibinfo {pages}
  {043009} (\bibinfo {year} {2022})},\ \Eprint
  {http://arxiv.org/abs/2201.02252} {arXiv:2201.02252 [astro-ph.HE]}
  \BibitemShut {NoStop}%
\bibitem [{\citenamefont {Will}(2014)}]{Will:2014kxa}%
  \BibitemOpen
  \bibfield  {author} {\bibinfo {author} {\bibfnamefont {C.~M.}\ \bibnamefont
  {Will}},\ }\href {\doibase 10.12942/lrr-2014-4} {\bibfield  {journal}
  {\bibinfo  {journal} {Living Rev. Rel.}\ }\textbf {\bibinfo {volume} {17}},\
  \bibinfo {pages} {4} (\bibinfo {year} {2014})},\ \Eprint
  {http://arxiv.org/abs/1403.7377} {arXiv:1403.7377 [gr-qc]} \BibitemShut
  {NoStop}%
\bibitem [{\citenamefont {Sathyaprakash}\ \emph {et~al.}(2019)\citenamefont
  {Sathyaprakash} \emph {et~al.}}]{Sathyaprakash:2019yqt}%
  \BibitemOpen
  \bibfield  {author} {\bibinfo {author} {\bibfnamefont {B.~S.}\ \bibnamefont
  {Sathyaprakash}} \emph {et~al.},\ }\href@noop {} {\  (\bibinfo {year}
  {2019})},\ \Eprint {http://arxiv.org/abs/1903.09221} {arXiv:1903.09221
  [astro-ph.HE]} \BibitemShut {NoStop}%
\bibitem [{\citenamefont {Berti}\ \emph {et~al.}(2016)\citenamefont {Berti},
  \citenamefont {Sesana}, \citenamefont {Barausse}, \citenamefont {Cardoso},\
  and\ \citenamefont {Belczynski}}]{Berti:2016lat}%
  \BibitemOpen
  \bibfield  {author} {\bibinfo {author} {\bibfnamefont {E.}~\bibnamefont
  {Berti}}, \bibinfo {author} {\bibfnamefont {A.}~\bibnamefont {Sesana}},
  \bibinfo {author} {\bibfnamefont {E.}~\bibnamefont {Barausse}}, \bibinfo
  {author} {\bibfnamefont {V.}~\bibnamefont {Cardoso}}, \ and\ \bibinfo
  {author} {\bibfnamefont {K.}~\bibnamefont {Belczynski}},\ }\href {\doibase
  10.1103/PhysRevLett.117.101102} {\bibfield  {journal} {\bibinfo  {journal}
  {Phys. Rev. Lett.}\ }\textbf {\bibinfo {volume} {117}},\ \bibinfo {pages}
  {101102} (\bibinfo {year} {2016})},\ \Eprint
  {http://arxiv.org/abs/1605.09286} {arXiv:1605.09286 [gr-qc]} \BibitemShut
  {NoStop}%
\bibitem [{\citenamefont {Dreyer}\ \emph {et~al.}(2004)\citenamefont {Dreyer},
  \citenamefont {Kelly}, \citenamefont {Krishnan}, \citenamefont {Finn},
  \citenamefont {Garrison},\ and\ \citenamefont
  {Lopez-Aleman}}]{Dreyer:2003bv}%
  \BibitemOpen
  \bibfield  {author} {\bibinfo {author} {\bibfnamefont {O.}~\bibnamefont
  {Dreyer}}, \bibinfo {author} {\bibfnamefont {B.~J.}\ \bibnamefont {Kelly}},
  \bibinfo {author} {\bibfnamefont {B.}~\bibnamefont {Krishnan}}, \bibinfo
  {author} {\bibfnamefont {L.~S.}\ \bibnamefont {Finn}}, \bibinfo {author}
  {\bibfnamefont {D.}~\bibnamefont {Garrison}}, \ and\ \bibinfo {author}
  {\bibfnamefont {R.}~\bibnamefont {Lopez-Aleman}},\ }\href {\doibase
  10.1088/0264-9381/21/4/003} {\bibfield  {journal} {\bibinfo  {journal}
  {Class. Quant. Grav.}\ }\textbf {\bibinfo {volume} {21}},\ \bibinfo {pages}
  {787} (\bibinfo {year} {2004})},\ \Eprint
  {http://arxiv.org/abs/gr-qc/0309007} {arXiv:gr-qc/0309007} \BibitemShut
  {NoStop}%
\bibitem [{\citenamefont {Berti}\ \emph
  {et~al.}(2018{\natexlab{a}})\citenamefont {Berti}, \citenamefont {Yagi},\
  and\ \citenamefont {Yunes}}]{Berti:2018cxi}%
  \BibitemOpen
  \bibfield  {author} {\bibinfo {author} {\bibfnamefont {E.}~\bibnamefont
  {Berti}}, \bibinfo {author} {\bibfnamefont {K.}~\bibnamefont {Yagi}}, \ and\
  \bibinfo {author} {\bibfnamefont {N.}~\bibnamefont {Yunes}},\ }\href
  {\doibase 10.1007/s10714-018-2362-8} {\bibfield  {journal} {\bibinfo
  {journal} {Gen. Rel. Grav.}\ }\textbf {\bibinfo {volume} {50}},\ \bibinfo
  {pages} {46} (\bibinfo {year} {2018}{\natexlab{a}})},\ \Eprint
  {http://arxiv.org/abs/1801.03208} {arXiv:1801.03208 [gr-qc]} \BibitemShut
  {NoStop}%
\bibitem [{\citenamefont {Berti}\ \emph
  {et~al.}(2018{\natexlab{b}})\citenamefont {Berti}, \citenamefont {Yagi},
  \citenamefont {Yang},\ and\ \citenamefont {Yunes}}]{Berti:2018vdi}%
  \BibitemOpen
  \bibfield  {author} {\bibinfo {author} {\bibfnamefont {E.}~\bibnamefont
  {Berti}}, \bibinfo {author} {\bibfnamefont {K.}~\bibnamefont {Yagi}},
  \bibinfo {author} {\bibfnamefont {H.}~\bibnamefont {Yang}}, \ and\ \bibinfo
  {author} {\bibfnamefont {N.}~\bibnamefont {Yunes}},\ }\href {\doibase
  10.1007/s10714-018-2372-6} {\bibfield  {journal} {\bibinfo  {journal} {Gen.
  Rel. Grav.}\ }\textbf {\bibinfo {volume} {50}},\ \bibinfo {pages} {49}
  (\bibinfo {year} {2018}{\natexlab{b}})},\ \Eprint
  {http://arxiv.org/abs/1801.03587} {arXiv:1801.03587 [gr-qc]} \BibitemShut
  {NoStop}%
\bibitem [{\citenamefont {Barack}\ \emph {et~al.}(2019)\citenamefont {Barack}
  \emph {et~al.}}]{Barack:2018yly}%
  \BibitemOpen
  \bibfield  {author} {\bibinfo {author} {\bibfnamefont {L.}~\bibnamefont
  {Barack}} \emph {et~al.},\ }\href {\doibase 10.1088/1361-6382/ab0587}
  {\bibfield  {journal} {\bibinfo  {journal} {Class. Quant. Grav.}\ }\textbf
  {\bibinfo {volume} {36}},\ \bibinfo {pages} {143001} (\bibinfo {year}
  {2019})},\ \Eprint {http://arxiv.org/abs/1806.05195} {arXiv:1806.05195
  [gr-qc]} \BibitemShut {NoStop}%
\bibitem [{\citenamefont {Sathyaprakash}\ and\ \citenamefont
  {Schutz}(2009)}]{Sathyaprakash:2009xs}%
  \BibitemOpen
  \bibfield  {author} {\bibinfo {author} {\bibfnamefont {B.~S.}\ \bibnamefont
  {Sathyaprakash}}\ and\ \bibinfo {author} {\bibfnamefont {B.~F.}\ \bibnamefont
  {Schutz}},\ }\href {\doibase 10.12942/lrr-2009-2} {\bibfield  {journal}
  {\bibinfo  {journal} {Living Rev. Rel.}\ }\textbf {\bibinfo {volume} {12}},\
  \bibinfo {pages} {2} (\bibinfo {year} {2009})},\ \Eprint
  {http://arxiv.org/abs/0903.0338} {arXiv:0903.0338 [gr-qc]} \BibitemShut
  {NoStop}%
\bibitem [{\citenamefont {Agullo}\ \emph {et~al.}(2021)\citenamefont {Agullo},
  \citenamefont {Cardoso}, \citenamefont {del Rio}, \citenamefont {Maggiore},\
  and\ \citenamefont {Pullin}}]{Agullo:2021}%
  \BibitemOpen
  \bibfield  {author} {\bibinfo {author} {\bibfnamefont {I.}~\bibnamefont
  {Agullo}}, \bibinfo {author} {\bibfnamefont {V.}~\bibnamefont {Cardoso}},
  \bibinfo {author} {\bibfnamefont {A.}~\bibnamefont {del Rio}}, \bibinfo
  {author} {\bibfnamefont {M.}~\bibnamefont {Maggiore}}, \ and\ \bibinfo
  {author} {\bibfnamefont {J.}~\bibnamefont {Pullin}},\ }\href@noop {}
  {\bibfield  {journal} {\bibinfo  {journal} {Phys. Rev. Lett.}\ }\textbf
  {\bibinfo {volume} {126}},\ \bibinfo {pages} {041302} (\bibinfo {year}
  {2021})},\ \Eprint {http://arxiv.org/abs/2007.13761} {arXiv:2007.13761
  [gr-qc]} \BibitemShut {NoStop}%
\bibitem [{\citenamefont {Cardoso}\ \emph {et~al.}(2019)\citenamefont
  {Cardoso}, \citenamefont {Foit},\ and\ \citenamefont
  {Kleban}}]{Cardoso:2019}%
  \BibitemOpen
  \bibfield  {author} {\bibinfo {author} {\bibfnamefont {V.}~\bibnamefont
  {Cardoso}}, \bibinfo {author} {\bibfnamefont {V.~F.}\ \bibnamefont {Foit}}, \
  and\ \bibinfo {author} {\bibfnamefont {M.}~\bibnamefont {Kleban}},\
  }\href@noop {} {\bibfield  {journal} {\bibinfo  {journal} {J. Cosmol.
  Astropart. Phys.}\ }\textbf {\bibinfo {volume} {08}},\ \bibinfo {pages} {006}
  (\bibinfo {year} {2019})},\ \Eprint {http://arxiv.org/abs/1902.10164}
  {arXiv:1902.10164 [hep-th]} \BibitemShut {NoStop}%
\bibitem [{\citenamefont {Cardoso}\ and\ \citenamefont
  {Pani}(2019)}]{Cardoso:2019rvt}%
  \BibitemOpen
  \bibfield  {author} {\bibinfo {author} {\bibfnamefont {V.}~\bibnamefont
  {Cardoso}}\ and\ \bibinfo {author} {\bibfnamefont {P.}~\bibnamefont {Pani}},\
  }\href {\doibase 10.1007/s41114-019-0020-4} {\bibfield  {journal} {\bibinfo
  {journal} {Living Rev. Rel.}\ }\textbf {\bibinfo {volume} {22}},\ \bibinfo
  {pages} {4} (\bibinfo {year} {2019})},\ \Eprint
  {http://arxiv.org/abs/1904.05363} {arXiv:1904.05363 [gr-qc]} \BibitemShut
  {NoStop}%
\bibitem [{\citenamefont {Pani}\ \emph {et~al.}(2009)\citenamefont {Pani},
  \citenamefont {Berti}, \citenamefont {Cardoso}, \citenamefont {Chen},\ and\
  \citenamefont {Norte}}]{Pani:2009ss}%
  \BibitemOpen
  \bibfield  {author} {\bibinfo {author} {\bibfnamefont {P.}~\bibnamefont
  {Pani}}, \bibinfo {author} {\bibfnamefont {E.}~\bibnamefont {Berti}},
  \bibinfo {author} {\bibfnamefont {V.}~\bibnamefont {Cardoso}}, \bibinfo
  {author} {\bibfnamefont {Y.}~\bibnamefont {Chen}}, \ and\ \bibinfo {author}
  {\bibfnamefont {R.}~\bibnamefont {Norte}},\ }\href {\doibase
  10.1103/PhysRevD.80.124047} {\bibfield  {journal} {\bibinfo  {journal} {Phys.
  Rev. D}\ }\textbf {\bibinfo {volume} {80}},\ \bibinfo {pages} {124047}
  (\bibinfo {year} {2009})},\ \Eprint {http://arxiv.org/abs/0909.0287}
  {arXiv:0909.0287 [gr-qc]} \BibitemShut {NoStop}%
\bibitem [{\citenamefont {Poisson}(2004)}]{Poisson:2004cw}%
  \BibitemOpen
  \bibfield  {author} {\bibinfo {author} {\bibfnamefont {E.}~\bibnamefont
  {Poisson}},\ }\href {\doibase 10.1103/PhysRevD.70.084044} {\bibfield
  {journal} {\bibinfo  {journal} {Phys. Rev.}\ }\textbf {\bibinfo {volume}
  {D70}},\ \bibinfo {pages} {084044} (\bibinfo {year} {2004})},\ \Eprint
  {http://arxiv.org/abs/gr-qc/0407050} {arXiv:gr-qc/0407050 [gr-qc]}
  \BibitemShut {NoStop}%
\bibitem [{\citenamefont {Alvi}(2001)}]{Alvi:2001mx}%
  \BibitemOpen
  \bibfield  {author} {\bibinfo {author} {\bibfnamefont {K.}~\bibnamefont
  {Alvi}},\ }\href {\doibase 10.1103/PhysRevD.64.104020} {\bibfield  {journal}
  {\bibinfo  {journal} {Phys. Rev. D}\ }\textbf {\bibinfo {volume} {64}},\
  \bibinfo {pages} {104020} (\bibinfo {year} {2001})},\ \Eprint
  {http://arxiv.org/abs/gr-qc/0107080} {arXiv:gr-qc/0107080} \BibitemShut
  {NoStop}%
\bibitem [{\citenamefont {Chakraborty}\ \emph {et~al.}(2021)\citenamefont
  {Chakraborty}, \citenamefont {Datta},\ and\ \citenamefont
  {Sau}}]{Chakraborty:2021gdf}%
  \BibitemOpen
  \bibfield  {author} {\bibinfo {author} {\bibfnamefont {S.}~\bibnamefont
  {Chakraborty}}, \bibinfo {author} {\bibfnamefont {S.}~\bibnamefont {Datta}},
  \ and\ \bibinfo {author} {\bibfnamefont {S.}~\bibnamefont {Sau}},\ }\href
  {\doibase 10.1103/PhysRevD.104.104001} {\bibfield  {journal} {\bibinfo
  {journal} {Phys. Rev. D}\ }\textbf {\bibinfo {volume} {104}},\ \bibinfo
  {pages} {104001} (\bibinfo {year} {2021})},\ \Eprint
  {http://arxiv.org/abs/2103.12430} {arXiv:2103.12430 [gr-qc]} \BibitemShut
  {NoStop}%
\bibitem [{\citenamefont {Datta}(2020)}]{Datta:2020rvo}%
  \BibitemOpen
  \bibfield  {author} {\bibinfo {author} {\bibfnamefont {S.}~\bibnamefont
  {Datta}},\ }\href {\doibase 10.1103/PhysRevD.102.064040} {\bibfield
  {journal} {\bibinfo  {journal} {Phys. Rev. D}\ }\textbf {\bibinfo {volume}
  {102}},\ \bibinfo {pages} {064040} (\bibinfo {year} {2020})},\ \Eprint
  {http://arxiv.org/abs/2002.04480} {arXiv:2002.04480 [gr-qc]} \BibitemShut
  {NoStop}%
\bibitem [{\citenamefont {Datta}\ \emph {et~al.}(2020)\citenamefont {Datta},
  \citenamefont {Brito}, \citenamefont {Bose}, \citenamefont {Pani},\ and\
  \citenamefont {Hughes}}]{Datta:2019epe}%
  \BibitemOpen
  \bibfield  {author} {\bibinfo {author} {\bibfnamefont {S.}~\bibnamefont
  {Datta}}, \bibinfo {author} {\bibfnamefont {R.}~\bibnamefont {Brito}},
  \bibinfo {author} {\bibfnamefont {S.}~\bibnamefont {Bose}}, \bibinfo {author}
  {\bibfnamefont {P.}~\bibnamefont {Pani}}, \ and\ \bibinfo {author}
  {\bibfnamefont {S.~A.}\ \bibnamefont {Hughes}},\ }\href {\doibase
  10.1103/PhysRevD.101.044004} {\bibfield  {journal} {\bibinfo  {journal}
  {Phys. Rev. D}\ }\textbf {\bibinfo {volume} {101}},\ \bibinfo {pages}
  {044004} (\bibinfo {year} {2020})},\ \Eprint
  {http://arxiv.org/abs/1910.07841} {arXiv:1910.07841 [gr-qc]} \BibitemShut
  {NoStop}%
\bibitem [{\citenamefont {Chatziioannou}\ \emph {et~al.}(2013)\citenamefont
  {Chatziioannou}, \citenamefont {Poisson},\ and\ \citenamefont
  {Yunes}}]{Chatziioannou:2012gq}%
  \BibitemOpen
  \bibfield  {author} {\bibinfo {author} {\bibfnamefont {K.}~\bibnamefont
  {Chatziioannou}}, \bibinfo {author} {\bibfnamefont {E.}~\bibnamefont
  {Poisson}}, \ and\ \bibinfo {author} {\bibfnamefont {N.}~\bibnamefont
  {Yunes}},\ }\href {\doibase 10.1103/PhysRevD.87.044022} {\bibfield  {journal}
  {\bibinfo  {journal} {Phys. Rev. D}\ }\textbf {\bibinfo {volume} {87}},\
  \bibinfo {pages} {044022} (\bibinfo {year} {2013})},\ \Eprint
  {http://arxiv.org/abs/1211.1686} {arXiv:1211.1686 [gr-qc]} \BibitemShut
  {NoStop}%
\bibitem [{\citenamefont {Saketh}\ \emph {et~al.}(2023)\citenamefont {Saketh},
  \citenamefont {Steinhoff}, \citenamefont {Vines},\ and\ \citenamefont
  {Buonanno}}]{Saketh:2022xjb}%
  \BibitemOpen
  \bibfield  {author} {\bibinfo {author} {\bibfnamefont {M.~V.~S.}\
  \bibnamefont {Saketh}}, \bibinfo {author} {\bibfnamefont {J.}~\bibnamefont
  {Steinhoff}}, \bibinfo {author} {\bibfnamefont {J.}~\bibnamefont {Vines}}, \
  and\ \bibinfo {author} {\bibfnamefont {A.}~\bibnamefont {Buonanno}},\ }\href
  {\doibase 10.1103/PhysRevD.107.084006} {\bibfield  {journal} {\bibinfo
  {journal} {Phys. Rev. D}\ }\textbf {\bibinfo {volume} {107}},\ \bibinfo
  {pages} {084006} (\bibinfo {year} {2023})},\ \Eprint
  {http://arxiv.org/abs/2212.13095} {arXiv:2212.13095 [gr-qc]} \BibitemShut
  {NoStop}%
\bibitem [{\citenamefont {Chakraborty}\ \emph
  {et~al.}(2024{\natexlab{a}})\citenamefont {Chakraborty}, \citenamefont
  {Comp\`ere},\ and\ \citenamefont {Machet}}]{Chakraborty:2024gcr}%
  \BibitemOpen
  \bibfield  {author} {\bibinfo {author} {\bibfnamefont {S.}~\bibnamefont
  {Chakraborty}}, \bibinfo {author} {\bibfnamefont {G.}~\bibnamefont
  {Comp\`ere}}, \ and\ \bibinfo {author} {\bibfnamefont {L.}~\bibnamefont
  {Machet}},\ }\href@noop {} {\  (\bibinfo {year} {2024}{\natexlab{a}})},\
  \Eprint {http://arxiv.org/abs/2412.14831} {arXiv:2412.14831 [gr-qc]}
  \BibitemShut {NoStop}%
\bibitem [{\citenamefont {Chakraborty}\ \emph
  {et~al.}(2024{\natexlab{b}})\citenamefont {Chakraborty}, \citenamefont
  {Maggio}, \citenamefont {Silvestrini},\ and\ \citenamefont
  {Pani}}]{Chakraborty:2023zed}%
  \BibitemOpen
  \bibfield  {author} {\bibinfo {author} {\bibfnamefont {S.}~\bibnamefont
  {Chakraborty}}, \bibinfo {author} {\bibfnamefont {E.}~\bibnamefont {Maggio}},
  \bibinfo {author} {\bibfnamefont {M.}~\bibnamefont {Silvestrini}}, \ and\
  \bibinfo {author} {\bibfnamefont {P.}~\bibnamefont {Pani}},\ }\href {\doibase
  10.1103/PhysRevD.110.084042} {\bibfield  {journal} {\bibinfo  {journal}
  {Phys. Rev. D}\ }\textbf {\bibinfo {volume} {110}},\ \bibinfo {pages}
  {084042} (\bibinfo {year} {2024}{\natexlab{b}})},\ \Eprint
  {http://arxiv.org/abs/2310.06023} {arXiv:2310.06023 [gr-qc]} \BibitemShut
  {NoStop}%
\bibitem [{\citenamefont {Cardoso}\ \emph {et~al.}(2017)\citenamefont
  {Cardoso}, \citenamefont {Franzin}, \citenamefont {Maselli}, \citenamefont
  {Pani},\ and\ \citenamefont {Raposo}}]{Cardoso:2017cfl}%
  \BibitemOpen
  \bibfield  {author} {\bibinfo {author} {\bibfnamefont {V.}~\bibnamefont
  {Cardoso}}, \bibinfo {author} {\bibfnamefont {E.}~\bibnamefont {Franzin}},
  \bibinfo {author} {\bibfnamefont {A.}~\bibnamefont {Maselli}}, \bibinfo
  {author} {\bibfnamefont {P.}~\bibnamefont {Pani}}, \ and\ \bibinfo {author}
  {\bibfnamefont {G.}~\bibnamefont {Raposo}},\ }\href {\doibase
  10.1103/PhysRevD.95.084014} {\bibfield  {journal} {\bibinfo  {journal} {Phys.
  Rev. D}\ }\textbf {\bibinfo {volume} {95}},\ \bibinfo {pages} {084014}
  (\bibinfo {year} {2017})},\ \bibinfo {note} {[Addendum: Phys.Rev.D 95, 089901
  (2017)]},\ \Eprint {http://arxiv.org/abs/1701.01116} {arXiv:1701.01116
  [gr-qc]} \BibitemShut {NoStop}%
\bibitem [{\citenamefont {Creci}\ \emph {et~al.}(2021)\citenamefont {Creci},
  \citenamefont {Hinderer},\ and\ \citenamefont {Steinhoff}}]{Creci:2021rkz}%
  \BibitemOpen
  \bibfield  {author} {\bibinfo {author} {\bibfnamefont {G.}~\bibnamefont
  {Creci}}, \bibinfo {author} {\bibfnamefont {T.}~\bibnamefont {Hinderer}}, \
  and\ \bibinfo {author} {\bibfnamefont {J.}~\bibnamefont {Steinhoff}},\ }\href
  {\doibase 10.1103/PhysRevD.104.124061} {\bibfield  {journal} {\bibinfo
  {journal} {Phys. Rev. D}\ }\textbf {\bibinfo {volume} {104}},\ \bibinfo
  {pages} {124061} (\bibinfo {year} {2021})},\ \bibinfo {note} {[Erratum:
  Phys.Rev.D 105, 109902 (2022)]},\ \Eprint {http://arxiv.org/abs/2108.03385}
  {arXiv:2108.03385 [gr-qc]} \BibitemShut {NoStop}%
\bibitem [{\citenamefont {Chakravarti}\ \emph {et~al.}(2019)\citenamefont
  {Chakravarti}, \citenamefont {Chakraborty}, \citenamefont {Bose},\ and\
  \citenamefont {SenGupta}}]{Chakravarti:2018vlt}%
  \BibitemOpen
  \bibfield  {author} {\bibinfo {author} {\bibfnamefont {K.}~\bibnamefont
  {Chakravarti}}, \bibinfo {author} {\bibfnamefont {S.}~\bibnamefont
  {Chakraborty}}, \bibinfo {author} {\bibfnamefont {S.}~\bibnamefont {Bose}}, \
  and\ \bibinfo {author} {\bibfnamefont {S.}~\bibnamefont {SenGupta}},\ }\href
  {\doibase 10.1103/PhysRevD.99.024036} {\bibfield  {journal} {\bibinfo
  {journal} {Phys. Rev. D}\ }\textbf {\bibinfo {volume} {99}},\ \bibinfo
  {pages} {024036} (\bibinfo {year} {2019})},\ \Eprint
  {http://arxiv.org/abs/1811.11364} {arXiv:1811.11364 [gr-qc]} \BibitemShut
  {NoStop}%
\bibitem [{\citenamefont {Binnington}\ and\ \citenamefont
  {Poisson}(2009)}]{Binnington:2009bb}%
  \BibitemOpen
  \bibfield  {author} {\bibinfo {author} {\bibfnamefont {T.}~\bibnamefont
  {Binnington}}\ and\ \bibinfo {author} {\bibfnamefont {E.}~\bibnamefont
  {Poisson}},\ }\href {\doibase 10.1103/PhysRevD.80.084018} {\bibfield
  {journal} {\bibinfo  {journal} {Phys. Rev.}\ }\textbf {\bibinfo {volume}
  {D80}},\ \bibinfo {pages} {084018} (\bibinfo {year} {2009})},\ \Eprint
  {http://arxiv.org/abs/0906.1366} {arXiv:0906.1366 [gr-qc]} \BibitemShut
  {NoStop}%
\bibitem [{\citenamefont {Nagar}\ and\ \citenamefont
  {Akcay}(2012)}]{Nagar:2011aa}%
  \BibitemOpen
  \bibfield  {author} {\bibinfo {author} {\bibfnamefont {A.}~\bibnamefont
  {Nagar}}\ and\ \bibinfo {author} {\bibfnamefont {S.}~\bibnamefont {Akcay}},\
  }\href {\doibase 10.1103/PhysRevD.85.044025} {\bibfield  {journal} {\bibinfo
  {journal} {Phys. Rev.}\ }\textbf {\bibinfo {volume} {D85}},\ \bibinfo {pages}
  {044025} (\bibinfo {year} {2012})},\ \Eprint {http://arxiv.org/abs/1112.2840}
  {arXiv:1112.2840 [gr-qc]} \BibitemShut {NoStop}%
\bibitem [{\citenamefont {Rosato}\ \emph {et~al.}(2025)\citenamefont {Rosato},
  \citenamefont {Biswas}, \citenamefont {Chakraborty},\ and\ \citenamefont
  {Pani}}]{Rosato:2025byu}%
  \BibitemOpen
  \bibfield  {author} {\bibinfo {author} {\bibfnamefont {R.~F.}\ \bibnamefont
  {Rosato}}, \bibinfo {author} {\bibfnamefont {S.}~\bibnamefont {Biswas}},
  \bibinfo {author} {\bibfnamefont {S.}~\bibnamefont {Chakraborty}}, \ and\
  \bibinfo {author} {\bibfnamefont {P.}~\bibnamefont {Pani}},\ }\href {\doibase
  10.1103/PhysRevD.111.084051} {\bibfield  {journal} {\bibinfo  {journal}
  {Phys. Rev. D}\ }\textbf {\bibinfo {volume} {111}},\ \bibinfo {pages}
  {084051} (\bibinfo {year} {2025})},\ \Eprint
  {http://arxiv.org/abs/2501.16433} {arXiv:2501.16433 [gr-qc]} \BibitemShut
  {NoStop}%
\bibitem [{\citenamefont {Abedi}\ \emph {et~al.}(2020)\citenamefont {Abedi},
  \citenamefont {Afshordi}, \citenamefont {Oshita},\ and\ \citenamefont
  {Wang}}]{Abedi:2020ujo}%
  \BibitemOpen
  \bibfield  {author} {\bibinfo {author} {\bibfnamefont {J.}~\bibnamefont
  {Abedi}}, \bibinfo {author} {\bibfnamefont {N.}~\bibnamefont {Afshordi}},
  \bibinfo {author} {\bibfnamefont {N.}~\bibnamefont {Oshita}}, \ and\ \bibinfo
  {author} {\bibfnamefont {Q.}~\bibnamefont {Wang}},\ }\href {\doibase
  10.3390/universe6030043} {\bibfield  {journal} {\bibinfo  {journal}
  {Universe}\ }\textbf {\bibinfo {volume} {6}},\ \bibinfo {pages} {43}
  (\bibinfo {year} {2020})},\ \Eprint {http://arxiv.org/abs/2001.09553}
  {arXiv:2001.09553 [gr-qc]} \BibitemShut {NoStop}%
\bibitem [{\citenamefont {Wang}\ \emph {et~al.}(2020)\citenamefont {Wang},
  \citenamefont {Oshita},\ and\ \citenamefont {Afshordi}}]{Wang:2019rcf}%
  \BibitemOpen
  \bibfield  {author} {\bibinfo {author} {\bibfnamefont {Q.}~\bibnamefont
  {Wang}}, \bibinfo {author} {\bibfnamefont {N.}~\bibnamefont {Oshita}}, \ and\
  \bibinfo {author} {\bibfnamefont {N.}~\bibnamefont {Afshordi}},\ }\href
  {\doibase 10.1103/PhysRevD.101.024031} {\bibfield  {journal} {\bibinfo
  {journal} {Phys. Rev. D}\ }\textbf {\bibinfo {volume} {101}},\ \bibinfo
  {pages} {024031} (\bibinfo {year} {2020})},\ \Eprint
  {http://arxiv.org/abs/1905.00446} {arXiv:1905.00446 [gr-qc]} \BibitemShut
  {NoStop}%
\bibitem [{\citenamefont {Mark}\ \emph {et~al.}(2017)\citenamefont {Mark},
  \citenamefont {Zimmerman}, \citenamefont {Du},\ and\ \citenamefont
  {Chen}}]{Mark:2017dnq}%
  \BibitemOpen
  \bibfield  {author} {\bibinfo {author} {\bibfnamefont {Z.}~\bibnamefont
  {Mark}}, \bibinfo {author} {\bibfnamefont {A.}~\bibnamefont {Zimmerman}},
  \bibinfo {author} {\bibfnamefont {S.~M.}\ \bibnamefont {Du}}, \ and\ \bibinfo
  {author} {\bibfnamefont {Y.}~\bibnamefont {Chen}},\ }\href {\doibase
  10.1103/PhysRevD.96.084002} {\bibfield  {journal} {\bibinfo  {journal} {Phys.
  Rev.}\ }\textbf {\bibinfo {volume} {D96}},\ \bibinfo {pages} {084002}
  (\bibinfo {year} {2017})},\ \Eprint {http://arxiv.org/abs/1706.06155}
  {arXiv:1706.06155 [gr-qc]} \BibitemShut {NoStop}%
\bibitem [{\citenamefont {Pani}\ and\ \citenamefont
  {Ferrari}(2018)}]{Pani:2018flj}%
  \BibitemOpen
  \bibfield  {author} {\bibinfo {author} {\bibfnamefont {P.}~\bibnamefont
  {Pani}}\ and\ \bibinfo {author} {\bibfnamefont {V.}~\bibnamefont {Ferrari}},\
  }\href {\doibase 10.1088/1361-6382/aacb8f} {\bibfield  {journal} {\bibinfo
  {journal} {Class. Quant. Grav.}\ }\textbf {\bibinfo {volume} {35}},\ \bibinfo
  {pages} {15LT01} (\bibinfo {year} {2018})},\ \Eprint
  {http://arxiv.org/abs/1804.01444} {arXiv:1804.01444 [gr-qc]} \BibitemShut
  {NoStop}%
\bibitem [{\citenamefont {Konoplya}\ and\ \citenamefont
  {Zhidenko}(2011)}]{Konoplya:2011qq}%
  \BibitemOpen
  \bibfield  {author} {\bibinfo {author} {\bibfnamefont {R.~A.}\ \bibnamefont
  {Konoplya}}\ and\ \bibinfo {author} {\bibfnamefont {A.}~\bibnamefont
  {Zhidenko}},\ }\href {\doibase 10.1103/RevModPhys.83.793} {\bibfield
  {journal} {\bibinfo  {journal} {Rev. Mod. Phys.}\ }\textbf {\bibinfo {volume}
  {83}},\ \bibinfo {pages} {793} (\bibinfo {year} {2011})},\ \Eprint
  {http://arxiv.org/abs/1102.4014} {arXiv:1102.4014 [gr-qc]} \BibitemShut
  {NoStop}%
\bibitem [{\citenamefont {Kokkotas}\ and\ \citenamefont
  {Schmidt}(1999)}]{Kokkotas:1999bd}%
  \BibitemOpen
  \bibfield  {author} {\bibinfo {author} {\bibfnamefont {K.~D.}\ \bibnamefont
  {Kokkotas}}\ and\ \bibinfo {author} {\bibfnamefont {B.~G.}\ \bibnamefont
  {Schmidt}},\ }\href {\doibase 10.12942/lrr-1999-2} {\bibfield  {journal}
  {\bibinfo  {journal} {Living Rev. Rel.}\ }\textbf {\bibinfo {volume} {2}},\
  \bibinfo {pages} {2} (\bibinfo {year} {1999})},\ \Eprint
  {http://arxiv.org/abs/gr-qc/9909058} {arXiv:gr-qc/9909058 [gr-qc]}
  \BibitemShut {NoStop}%
\bibitem [{\citenamefont {Biswas}\ \emph {et~al.}(2022)\citenamefont {Biswas},
  \citenamefont {Rahman},\ and\ \citenamefont {Chakraborty}}]{Biswas:2022wah}%
  \BibitemOpen
  \bibfield  {author} {\bibinfo {author} {\bibfnamefont {S.}~\bibnamefont
  {Biswas}}, \bibinfo {author} {\bibfnamefont {M.}~\bibnamefont {Rahman}}, \
  and\ \bibinfo {author} {\bibfnamefont {S.}~\bibnamefont {Chakraborty}},\
  }\href {\doibase 10.1103/PhysRevD.106.124003} {\bibfield  {journal} {\bibinfo
   {journal} {Phys. Rev. D}\ }\textbf {\bibinfo {volume} {106}},\ \bibinfo
  {pages} {124003} (\bibinfo {year} {2022})},\ \Eprint
  {http://arxiv.org/abs/2205.14743} {arXiv:2205.14743 [gr-qc]} \BibitemShut
  {NoStop}%
\bibitem [{\citenamefont {Chakraborty}\ \emph {et~al.}(2022)\citenamefont
  {Chakraborty}, \citenamefont {Maggio}, \citenamefont {Mazumdar},\ and\
  \citenamefont {Pani}}]{Chakraborty:2022zlq}%
  \BibitemOpen
  \bibfield  {author} {\bibinfo {author} {\bibfnamefont {S.}~\bibnamefont
  {Chakraborty}}, \bibinfo {author} {\bibfnamefont {E.}~\bibnamefont {Maggio}},
  \bibinfo {author} {\bibfnamefont {A.}~\bibnamefont {Mazumdar}}, \ and\
  \bibinfo {author} {\bibfnamefont {P.}~\bibnamefont {Pani}},\ }\href {\doibase
  10.1103/PhysRevD.106.024041} {\bibfield  {journal} {\bibinfo  {journal}
  {Phys. Rev. D}\ }\textbf {\bibinfo {volume} {106}},\ \bibinfo {pages}
  {024041} (\bibinfo {year} {2022})},\ \Eprint
  {http://arxiv.org/abs/2202.09111} {arXiv:2202.09111 [gr-qc]} \BibitemShut
  {NoStop}%
\bibitem [{\citenamefont {Scheel}\ \emph {et~al.}(2025)\citenamefont {Scheel}
  \emph {et~al.}}]{Scheel:2025jct}%
  \BibitemOpen
  \bibfield  {author} {\bibinfo {author} {\bibfnamefont {M.~A.}\ \bibnamefont
  {Scheel}} \emph {et~al.},\ }\href@noop {} {\  (\bibinfo {year} {2025})},\
  \Eprint {http://arxiv.org/abs/2505.13378} {arXiv:2505.13378 [gr-qc]}
  \BibitemShut {NoStop}%
\bibitem [{\citenamefont {Paul}\ \emph {et~al.}(2025)\citenamefont {Paul},
  \citenamefont {Maurya}, \citenamefont {Henry}, \citenamefont {Sharma},
  \citenamefont {Satheesh}, \citenamefont {Divyajyoti}, \citenamefont {Kumar},\
  and\ \citenamefont {Mishra}}]{Paul:2024ujx}%
  \BibitemOpen
  \bibfield  {author} {\bibinfo {author} {\bibfnamefont {K.}~\bibnamefont
  {Paul}}, \bibinfo {author} {\bibfnamefont {A.}~\bibnamefont {Maurya}},
  \bibinfo {author} {\bibfnamefont {Q.}~\bibnamefont {Henry}}, \bibinfo
  {author} {\bibfnamefont {K.}~\bibnamefont {Sharma}}, \bibinfo {author}
  {\bibfnamefont {P.}~\bibnamefont {Satheesh}}, \bibinfo {author} {\bibnamefont
  {Divyajyoti}}, \bibinfo {author} {\bibfnamefont {P.}~\bibnamefont {Kumar}}, \
  and\ \bibinfo {author} {\bibfnamefont {C.~K.}\ \bibnamefont {Mishra}},\
  }\href {\doibase 10.1103/PhysRevD.111.084074} {\bibfield  {journal} {\bibinfo
   {journal} {Phys. Rev. D}\ }\textbf {\bibinfo {volume} {111}},\ \bibinfo
  {pages} {084074} (\bibinfo {year} {2025})},\ \Eprint
  {http://arxiv.org/abs/2409.13866} {arXiv:2409.13866 [gr-qc]} \BibitemShut
  {NoStop}%
\bibitem [{\citenamefont {Datta}\ and\ \citenamefont
  {Phukon}(2021)}]{Datta:2021row}%
  \BibitemOpen
  \bibfield  {author} {\bibinfo {author} {\bibfnamefont {S.}~\bibnamefont
  {Datta}}\ and\ \bibinfo {author} {\bibfnamefont {K.~S.}\ \bibnamefont
  {Phukon}},\ }\href {\doibase 10.1103/PhysRevD.104.124062} {\bibfield
  {journal} {\bibinfo  {journal} {Phys. Rev. D}\ }\textbf {\bibinfo {volume}
  {104}},\ \bibinfo {pages} {124062} (\bibinfo {year} {2021})},\ \Eprint
  {http://arxiv.org/abs/2105.11140} {arXiv:2105.11140 [gr-qc]} \BibitemShut
  {NoStop}%
\bibitem [{\citenamefont {Chakravarti}\ \emph {et~al.}(2021)\citenamefont
  {Chakravarti}, \citenamefont {Ghosh},\ and\ \citenamefont
  {Sarkar}}]{Chakravarti:2021jbv}%
  \BibitemOpen
  \bibfield  {author} {\bibinfo {author} {\bibfnamefont {K.}~\bibnamefont
  {Chakravarti}}, \bibinfo {author} {\bibfnamefont {R.}~\bibnamefont {Ghosh}},
  \ and\ \bibinfo {author} {\bibfnamefont {S.}~\bibnamefont {Sarkar}},\ }\href
  {\doibase 10.1103/PhysRevD.104.084049} {\bibfield  {journal} {\bibinfo
  {journal} {Phys. Rev. D}\ }\textbf {\bibinfo {volume} {104}},\ \bibinfo
  {pages} {084049} (\bibinfo {year} {2021})},\ \Eprint
  {http://arxiv.org/abs/2108.02444} {arXiv:2108.02444 [gr-qc]} \BibitemShut
  {NoStop}%
\bibitem [{\citenamefont {Chakravarti}\ \emph {et~al.}(2022)\citenamefont
  {Chakravarti}, \citenamefont {Ghosh},\ and\ \citenamefont
  {Sarkar}}]{Chakravarti:2021clm}%
  \BibitemOpen
  \bibfield  {author} {\bibinfo {author} {\bibfnamefont {K.}~\bibnamefont
  {Chakravarti}}, \bibinfo {author} {\bibfnamefont {R.}~\bibnamefont {Ghosh}},
  \ and\ \bibinfo {author} {\bibfnamefont {S.}~\bibnamefont {Sarkar}},\ }\href
  {\doibase 10.1103/PhysRevD.105.044046} {\bibfield  {journal} {\bibinfo
  {journal} {Phys. Rev. D}\ }\textbf {\bibinfo {volume} {105}},\ \bibinfo
  {pages} {044046} (\bibinfo {year} {2022})},\ \Eprint
  {http://arxiv.org/abs/2112.10109} {arXiv:2112.10109 [gr-qc]} \BibitemShut
  {NoStop}%
\bibitem [{\citenamefont {Krishnendu}\ \emph {et~al.}(2025)\citenamefont
  {Krishnendu}, \citenamefont {Ghosh}, \citenamefont {Datta}, \citenamefont
  {Chakraborty},\ and\ \citenamefont {Ajith}}]{Krishnendu:2025}%
  \BibitemOpen
  \bibfield  {author} {\bibinfo {author} {\bibfnamefont {N.}~\bibnamefont
  {Krishnendu}}, \bibinfo {author} {\bibfnamefont {R.}~\bibnamefont {Ghosh}},
  \bibinfo {author} {\bibfnamefont {S.}~\bibnamefont {Datta}}, \bibinfo
  {author} {\bibfnamefont {S.}~\bibnamefont {Chakraborty}}, \ and\ \bibinfo
  {author} {\bibfnamefont {P.}~\bibnamefont {Ajith}},\ }\href@noop {}
  {\bibfield  {journal} {\bibinfo  {journal} {Work in Progress}\ } (\bibinfo
  {year} {2025})}\BibitemShut {NoStop}%
\bibitem [{\citenamefont {Bekenstein}(1974)}]{Bekenstein:1974}%
  \BibitemOpen
  \bibfield  {author} {\bibinfo {author} {\bibfnamefont {J.~D.}\ \bibnamefont
  {Bekenstein}},\ }\href@noop {} {\bibfield  {journal} {\bibinfo  {journal}
  {Lett. Nuovo Cimento}\ }\textbf {\bibinfo {volume} {11}},\ \bibinfo {pages}
  {467} (\bibinfo {year} {1974})}\BibitemShut {NoStop}%
\bibitem [{\citenamefont {Maggiore}(2008)}]{Maggiore:2008}%
  \BibitemOpen
  \bibfield  {author} {\bibinfo {author} {\bibfnamefont {M.}~\bibnamefont
  {Maggiore}},\ }\href@noop {} {\bibfield  {journal} {\bibinfo  {journal}
  {Phys. Rev. Lett.}\ }\textbf {\bibinfo {volume} {100}},\ \bibinfo {pages}
  {141301} (\bibinfo {year} {2008})},\ \Eprint {http://arxiv.org/abs/0711.3145}
  {arXiv:0711.3145 [gr-qc]} \BibitemShut {NoStop}%
\bibitem [{\citenamefont {Chakraborty}\ and\ \citenamefont
  {Lochan}(2019)}]{Chakraborty:2017opo}%
  \BibitemOpen
  \bibfield  {author} {\bibinfo {author} {\bibfnamefont {S.}~\bibnamefont
  {Chakraborty}}\ and\ \bibinfo {author} {\bibfnamefont {K.}~\bibnamefont
  {Lochan}},\ }\href {\doibase 10.1016/j.physletb.2018.12.028} {\bibfield
  {journal} {\bibinfo  {journal} {Phys. Lett. B}\ }\textbf {\bibinfo {volume}
  {789}},\ \bibinfo {pages} {276} (\bibinfo {year} {2019})},\ \Eprint
  {http://arxiv.org/abs/1711.10660} {arXiv:1711.10660 [gr-qc]} \BibitemShut
  {NoStop}%
\bibitem [{\citenamefont {Lochan}\ and\ \citenamefont
  {Chakraborty}(2016)}]{Lochan:2015bha}%
  \BibitemOpen
  \bibfield  {author} {\bibinfo {author} {\bibfnamefont {K.}~\bibnamefont
  {Lochan}}\ and\ \bibinfo {author} {\bibfnamefont {S.}~\bibnamefont
  {Chakraborty}},\ }\href {\doibase 10.1016/j.physletb.2016.01.060} {\bibfield
  {journal} {\bibinfo  {journal} {Phys. Lett. B}\ }\textbf {\bibinfo {volume}
  {755}},\ \bibinfo {pages} {37} (\bibinfo {year} {2016})},\ \Eprint
  {http://arxiv.org/abs/1509.09010} {arXiv:1509.09010 [gr-qc]} \BibitemShut
  {NoStop}%
\bibitem [{\citenamefont {Hod}(1998)}]{Hod:1998}%
  \BibitemOpen
  \bibfield  {author} {\bibinfo {author} {\bibfnamefont {S.}~\bibnamefont
  {Hod}},\ }\href@noop {} {\bibfield  {journal} {\bibinfo  {journal} {Phys.
  Rev. Lett.}\ }\textbf {\bibinfo {volume} {81}},\ \bibinfo {pages} {4293}
  (\bibinfo {year} {1998})},\ \Eprint {http://arxiv.org/abs/gr-qc/9812002}
  {arXiv:gr-qc/9812002} \BibitemShut {NoStop}%
\bibitem [{\citenamefont {Mukhanov}(1986)}]{Mukhanov:1986}%
  \BibitemOpen
  \bibfield  {author} {\bibinfo {author} {\bibfnamefont {V.~F.}\ \bibnamefont
  {Mukhanov}},\ }\href@noop {} {\bibfield  {journal} {\bibinfo  {journal} {JETP
  Letters}\ }\textbf {\bibinfo {volume} {44}},\ \bibinfo {pages} {63} (\bibinfo
  {year} {1986})}\BibitemShut {NoStop}%
\bibitem [{\citenamefont {Barbero~G.}\ \emph {et~al.}(2009)\citenamefont
  {Barbero~G.}, \citenamefont {Lewandowski},\ and\ \citenamefont
  {Villase\~{n}or}}]{Barbero:2009}%
  \BibitemOpen
  \bibfield  {author} {\bibinfo {author} {\bibfnamefont {J.~F.}\ \bibnamefont
  {Barbero~G.}}, \bibinfo {author} {\bibfnamefont {J.}~\bibnamefont
  {Lewandowski}}, \ and\ \bibinfo {author} {\bibfnamefont {E.~J.~S.}\
  \bibnamefont {Villase\~{n}or}},\ }\href@noop {} {\bibfield  {journal}
  {\bibinfo  {journal} {Phys. Rev. D}\ }\textbf {\bibinfo {volume} {80}},\
  \bibinfo {pages} {044016} (\bibinfo {year} {2009})},\ \Eprint
  {http://arxiv.org/abs/0905.3465} {arXiv:0905.3465 [gr-qc]} \BibitemShut
  {NoStop}%
\bibitem [{\citenamefont
  {Padmanabhan}(1985{\natexlab{a}})}]{Padmanabhan:1985jq}%
  \BibitemOpen
  \bibfield  {author} {\bibinfo {author} {\bibfnamefont {T.}~\bibnamefont
  {Padmanabhan}},\ }\href {\doibase 10.1007/BF00760244} {\bibfield  {journal}
  {\bibinfo  {journal} {Gen. Rel. Grav.}\ }\textbf {\bibinfo {volume} {17}},\
  \bibinfo {pages} {215} (\bibinfo {year} {1985}{\natexlab{a}})}\BibitemShut
  {NoStop}%
\bibitem [{\citenamefont
  {Padmanabhan}(1985{\natexlab{b}})}]{Padmanabhan:1985jdl}%
  \BibitemOpen
  \bibfield  {author} {\bibinfo {author} {\bibfnamefont {T.}~\bibnamefont
  {Padmanabhan}},\ }\href {\doibase 10.1016/S0003-4916(85)80004-X} {\bibfield
  {journal} {\bibinfo  {journal} {Annals Phys.}\ }\textbf {\bibinfo {volume}
  {165}},\ \bibinfo {pages} {38} (\bibinfo {year}
  {1985}{\natexlab{b}})}\BibitemShut {NoStop}%
\bibitem [{\citenamefont {Garay}(1995)}]{Garay:1994en}%
  \BibitemOpen
  \bibfield  {author} {\bibinfo {author} {\bibfnamefont {L.~J.}\ \bibnamefont
  {Garay}},\ }\href {\doibase 10.1142/S0217751X95000085} {\bibfield  {journal}
  {\bibinfo  {journal} {Int. J. Mod. Phys. A}\ }\textbf {\bibinfo {volume}
  {10}},\ \bibinfo {pages} {145} (\bibinfo {year} {1995})},\ \Eprint
  {http://arxiv.org/abs/gr-qc/9403008} {arXiv:gr-qc/9403008} \BibitemShut
  {NoStop}%
\bibitem [{\citenamefont {Nicolini}(2022)}]{Nicolini:2022rlz}%
  \BibitemOpen
  \bibfield  {author} {\bibinfo {author} {\bibfnamefont {P.}~\bibnamefont
  {Nicolini}},\ }\href {\doibase 10.1007/s10714-022-02995-4} {\bibfield
  {journal} {\bibinfo  {journal} {Gen. Rel. Grav.}\ }\textbf {\bibinfo {volume}
  {54}},\ \bibinfo {pages} {106} (\bibinfo {year} {2022})},\ \Eprint
  {http://arxiv.org/abs/2208.05390} {arXiv:2208.05390 [hep-th]} \BibitemShut
  {NoStop}%
\bibitem [{\citenamefont {Padmanabhan}(1997)}]{Padmanabhan:1997}%
  \BibitemOpen
  \bibfield  {author} {\bibinfo {author} {\bibfnamefont {T.}~\bibnamefont
  {Padmanabhan}},\ }\href@noop {} {\bibfield  {journal} {\bibinfo  {journal}
  {Phys. Rev. Lett.}\ }\textbf {\bibinfo {volume} {78}} (\bibinfo {year}
  {1997})},\ \Eprint {http://arxiv.org/abs/hep-th/9608182}
  {arXiv:hep-th/9608182} \BibitemShut {NoStop}%
\bibitem [{\citenamefont {Fontanini}\ \emph {et~al.}(2006)\citenamefont
  {Fontanini}, \citenamefont {Spallucci},\ and\ \citenamefont
  {Padmanabhan}}]{Fontanini:2006}%
  \BibitemOpen
  \bibfield  {author} {\bibinfo {author} {\bibfnamefont {M.}~\bibnamefont
  {Fontanini}}, \bibinfo {author} {\bibfnamefont {E.}~\bibnamefont
  {Spallucci}}, \ and\ \bibinfo {author} {\bibfnamefont {T.}~\bibnamefont
  {Padmanabhan}},\ }\href@noop {} {\bibfield  {journal} {\bibinfo  {journal}
  {Phys. Lett. B}\ }\textbf {\bibinfo {volume} {633}},\ \bibinfo {pages} {627}
  (\bibinfo {year} {2006})},\ \Eprint {http://arxiv.org/abs/hep-th/0509090}
  {arXiv:hep-th/0509090} \BibitemShut {NoStop}%
\bibitem [{\citenamefont {Bishop}\ \emph {et~al.}(2023)\citenamefont {Bishop},
  \citenamefont {Contreras}, \citenamefont {Martin}, \citenamefont {Nicolini},\
  and\ \citenamefont {Singleton}}]{Bishop:2023hvz}%
  \BibitemOpen
  \bibfield  {author} {\bibinfo {author} {\bibfnamefont {M.}~\bibnamefont
  {Bishop}}, \bibinfo {author} {\bibfnamefont {J.}~\bibnamefont {Contreras}},
  \bibinfo {author} {\bibfnamefont {P.}~\bibnamefont {Martin}}, \bibinfo
  {author} {\bibfnamefont {P.}~\bibnamefont {Nicolini}}, \ and\ \bibinfo
  {author} {\bibfnamefont {D.}~\bibnamefont {Singleton}},\ }\href {\doibase
  10.1016/j.physletb.2023.138263} {\bibfield  {journal} {\bibinfo  {journal}
  {Phys. Lett. B}\ }\textbf {\bibinfo {volume} {847}},\ \bibinfo {pages}
  {138263} (\bibinfo {year} {2023})},\ \Eprint
  {http://arxiv.org/abs/2307.05367} {arXiv:2307.05367 [quant-ph]} \BibitemShut
  {NoStop}%
\bibitem [{\citenamefont {Kothawala}(2013)}]{KotE}%
  \BibitemOpen
  \bibfield  {author} {\bibinfo {author} {\bibfnamefont {D.}~\bibnamefont
  {Kothawala}},\ }\href {\doibase 10.1103/PhysRevD.88.104029} {\bibfield
  {journal} {\bibinfo  {journal} {Phys. Rev. D}\ }\textbf {\bibinfo {volume}
  {88}},\ \bibinfo {pages} {104029} (\bibinfo {year} {2013})},\ \Eprint
  {http://arxiv.org/abs/1307.5618} {arXiv:1307.5618 [gr-qc]} \BibitemShut
  {NoStop}%
\bibitem [{\citenamefont {Kothawala}\ and\ \citenamefont
  {Padmanabhan}(2014)}]{KotF}%
  \BibitemOpen
  \bibfield  {author} {\bibinfo {author} {\bibfnamefont {D.}~\bibnamefont
  {Kothawala}}\ and\ \bibinfo {author} {\bibfnamefont {T.}~\bibnamefont
  {Padmanabhan}},\ }\href {\doibase 10.1103/PhysRevD.90.124060} {\bibfield
  {journal} {\bibinfo  {journal} {Phys. Rev. D}\ }\textbf {\bibinfo {volume}
  {90}},\ \bibinfo {pages} {124060} (\bibinfo {year} {2014})},\ \Eprint
  {http://arxiv.org/abs/1405.4967} {arXiv:1405.4967 [gr-qc]} \BibitemShut
  {NoStop}%
\bibitem [{\citenamefont {Stargen}\ and\ \citenamefont
  {Kothawala}(2015)}]{KotI}%
  \BibitemOpen
  \bibfield  {author} {\bibinfo {author} {\bibfnamefont {D.~J.}\ \bibnamefont
  {Stargen}}\ and\ \bibinfo {author} {\bibfnamefont {D.}~\bibnamefont
  {Kothawala}},\ }\href {\doibase 10.1103/PhysRevD.92.024046} {\bibfield
  {journal} {\bibinfo  {journal} {Phys. Rev. D}\ }\textbf {\bibinfo {volume}
  {92}},\ \bibinfo {pages} {024046} (\bibinfo {year} {2015})},\ \Eprint
  {http://arxiv.org/abs/1503.03793} {arXiv:1503.03793 [gr-qc]} \BibitemShut
  {NoStop}%
\bibitem [{\citenamefont {Pesci}(2020)}]{Pesci:2020}%
  \BibitemOpen
  \bibfield  {author} {\bibinfo {author} {\bibfnamefont {A.}~\bibnamefont
  {Pesci}},\ }\href@noop {} {\bibfield  {journal} {\bibinfo  {journal} {Phys.
  Rev. D}\ }\textbf {\bibinfo {volume} {102}},\ \bibinfo {pages} {124057}
  (\bibinfo {year} {2020})},\ \Eprint {http://arxiv.org/abs/1911.04135}
  {arXiv:1911.04135 [gr-qc]} \BibitemShut {NoStop}%
\bibitem [{\citenamefont {Padmanabhan}(2020)}]{Padmanabhan:2020}%
  \BibitemOpen
  \bibfield  {author} {\bibinfo {author} {\bibfnamefont {T.}~\bibnamefont
  {Padmanabhan}},\ }\href@noop {} {\bibfield  {journal} {\bibinfo  {journal}
  {Int. J. Mod. Phys. D}\ }\textbf {\bibinfo {volume} {29}},\ \bibinfo {pages}
  {2030001} (\bibinfo {year} {2020})},\ \Eprint
  {http://arxiv.org/abs/1909.02015} {arXiv:1909.02015 [gr-qc]} \BibitemShut
  {NoStop}%
\bibitem [{\citenamefont {Padmanabhan}\ \emph {et~al.}(2016)\citenamefont
  {Padmanabhan}, \citenamefont {Chakraborty},\ and\ \citenamefont
  {Kothawala}}]{Padmanabhan:2016}%
  \BibitemOpen
  \bibfield  {author} {\bibinfo {author} {\bibfnamefont {T.}~\bibnamefont
  {Padmanabhan}}, \bibinfo {author} {\bibfnamefont {S.}~\bibnamefont
  {Chakraborty}}, \ and\ \bibinfo {author} {\bibfnamefont {D.}~\bibnamefont
  {Kothawala}},\ }\href@noop {} {\bibfield  {journal} {\bibinfo  {journal}
  {Gen. Rel. Grav.}\ }\textbf {\bibinfo {volume} {48}},\ \bibinfo {pages} {55}
  (\bibinfo {year} {2016})},\ \Eprint {http://arxiv.org/abs/1507.05669}
  {arXiv:1507.05669 [gr-qc]} \BibitemShut {NoStop}%
\bibitem [{\citenamefont {Chakraborty}\ \emph {et~al.}(2019)\citenamefont
  {Chakraborty}, \citenamefont {Kothawala},\ and\ \citenamefont
  {Pesci}}]{ChaD}%
  \BibitemOpen
  \bibfield  {author} {\bibinfo {author} {\bibfnamefont {S.}~\bibnamefont
  {Chakraborty}}, \bibinfo {author} {\bibfnamefont {D.}~\bibnamefont
  {Kothawala}}, \ and\ \bibinfo {author} {\bibfnamefont {A.}~\bibnamefont
  {Pesci}},\ }\href {\doibase 10.1016/j.physletb.2019.134877} {\bibfield
  {journal} {\bibinfo  {journal} {Phys. Lett. B}\ }\textbf {\bibinfo {volume}
  {797}},\ \bibinfo {pages} {134877} (\bibinfo {year} {2019})},\ \Eprint
  {http://arxiv.org/abs/1904.09053} {arXiv:1904.09053 [gr-qc]} \BibitemShut
  {NoStop}%
\bibitem [{\citenamefont {Chatziioannou}\ \emph {et~al.}(2016)\citenamefont
  {Chatziioannou}, \citenamefont {Poisson},\ and\ \citenamefont
  {Yunes}}]{Chatziioannou:2016kem}%
  \BibitemOpen
  \bibfield  {author} {\bibinfo {author} {\bibfnamefont {K.}~\bibnamefont
  {Chatziioannou}}, \bibinfo {author} {\bibfnamefont {E.}~\bibnamefont
  {Poisson}}, \ and\ \bibinfo {author} {\bibfnamefont {N.}~\bibnamefont
  {Yunes}},\ }\href {\doibase 10.1103/PhysRevD.94.084043} {\bibfield  {journal}
  {\bibinfo  {journal} {Phys. Rev. D}\ }\textbf {\bibinfo {volume} {94}},\
  \bibinfo {pages} {084043} (\bibinfo {year} {2016})},\ \Eprint
  {http://arxiv.org/abs/1608.02899} {arXiv:1608.02899 [gr-qc]} \BibitemShut
  {NoStop}%
\bibitem [{\citenamefont {Hartle}(1973)}]{Hartle1973}%
  \BibitemOpen
  \bibfield  {author} {\bibinfo {author} {\bibfnamefont {J.~B.}\ \bibnamefont
  {Hartle}},\ }\href {\doibase 10.1103/PhysRevD.8.1010} {\bibfield  {journal}
  {\bibinfo  {journal} {Phys. Rev. D}\ }\textbf {\bibinfo {volume} {8}},\
  \bibinfo {pages} {1010} (\bibinfo {year} {1973})}\BibitemShut {NoStop}%
\bibitem [{\citenamefont {Poisson}(2015)}]{Poisson2014}%
  \BibitemOpen
  \bibfield  {author} {\bibinfo {author} {\bibfnamefont {E.}~\bibnamefont
  {Poisson}},\ }\href {\doibase 10.1103/PhysRevD.91.044004} {\bibfield
  {journal} {\bibinfo  {journal} {Phys. Rev. D}\ }\textbf {\bibinfo {volume}
  {91}},\ \bibinfo {pages} {044004} (\bibinfo {year} {2015})},\ \Eprint
  {http://arxiv.org/abs/1411.4711} {arXiv:1411.4711 [gr-qc]} \BibitemShut
  {NoStop}%
\bibitem [{\citenamefont {Pesci}(2019)}]{QuantumMetricNull}%
  \BibitemOpen
  \bibfield  {author} {\bibinfo {author} {\bibfnamefont {A.}~\bibnamefont
  {Pesci}},\ }\href {\doibase 10.1088/1361-6382/ab0a40} {\bibfield  {journal}
  {\bibinfo  {journal} {Class. Quant. Grav.}\ }\textbf {\bibinfo {volume}
  {36}},\ \bibinfo {pages} {075009} (\bibinfo {year} {2019})},\ \bibinfo {note}
  {[Erratum: Class.Quant.Grav. 36, 229501 (2019)]},\ \Eprint
  {http://arxiv.org/abs/1812.01275} {arXiv:1812.01275 [gr-qc]} \BibitemShut
  {NoStop}%
\bibitem [{\citenamefont {{Van Vleck}}(1928)}]{vVl}%
  \BibitemOpen
  \bibfield  {author} {\bibinfo {author} {\bibfnamefont {J.~H.}\ \bibnamefont
  {{Van Vleck}}},\ }\href {\doibase 10.1073/pnas.14.2.178} {\bibfield
  {journal} {\bibinfo  {journal} {Proceedings of the National Academy of
  Science}\ }\textbf {\bibinfo {volume} {14}},\ \bibinfo {pages} {178}
  (\bibinfo {year} {1928})}\BibitemShut {NoStop}%
\bibitem [{\citenamefont {Morette}(1951)}]{Mor}%
  \BibitemOpen
  \bibfield  {author} {\bibinfo {author} {\bibfnamefont {C.}~\bibnamefont
  {Morette}},\ }\href {\doibase 10.1103/PhysRev.81.848} {\bibfield  {journal}
  {\bibinfo  {journal} {Phys. Rev.}\ }\textbf {\bibinfo {volume} {81}},\
  \bibinfo {pages} {848} (\bibinfo {year} {1951})}\BibitemShut {NoStop}%
\bibitem [{\citenamefont {DeWitt}\ and\ \citenamefont {Brehme}(1960)}]{DeWA}%
  \BibitemOpen
  \bibfield  {author} {\bibinfo {author} {\bibfnamefont {B.~S.}\ \bibnamefont
  {DeWitt}}\ and\ \bibinfo {author} {\bibfnamefont {R.~W.}\ \bibnamefont
  {Brehme}},\ }\href {\doibase 10.1016/0003-4916(60)90030-0} {\bibfield
  {journal} {\bibinfo  {journal} {Annals Phys.}\ }\textbf {\bibinfo {volume}
  {9}},\ \bibinfo {pages} {220} (\bibinfo {year} {1960})}\BibitemShut {NoStop}%
\bibitem [{\citenamefont {DeWitt}(1964)}]{DeWB}%
  \BibitemOpen
  \bibfield  {author} {\bibinfo {author} {\bibfnamefont {B.~S.}\ \bibnamefont
  {DeWitt}},\ }\href@noop {} {\bibfield  {journal} {\bibinfo  {journal} {Conf.
  Proc. C}\ }\textbf {\bibinfo {volume} {630701}},\ \bibinfo {pages} {585}
  (\bibinfo {year} {1964})}\BibitemShut {NoStop}%
\bibitem [{\citenamefont {Padmanabhan}(2015)}]{Pad06}%
  \BibitemOpen
  \bibfield  {author} {\bibinfo {author} {\bibfnamefont {T.}~\bibnamefont
  {Padmanabhan}},\ }\href {\doibase 10.3390/e17117420} {\bibfield  {journal}
  {\bibinfo  {journal} {Entropy}\ }\textbf {\bibinfo {volume} {17}},\ \bibinfo
  {pages} {7420} (\bibinfo {year} {2015})},\ \Eprint
  {http://arxiv.org/abs/1508.06286} {arXiv:1508.06286 [gr-qc]} \BibitemShut
  {NoStop}%
\bibitem [{\citenamefont {Krishnendu}\ and\ \citenamefont
  {Chakraborty}(2024)}]{Krishnendu:2024jkj}%
  \BibitemOpen
  \bibfield  {author} {\bibinfo {author} {\bibfnamefont {N.~V.}\ \bibnamefont
  {Krishnendu}}\ and\ \bibinfo {author} {\bibfnamefont {S.}~\bibnamefont
  {Chakraborty}},\ }\href {\doibase 10.1103/PhysRevD.109.124047} {\bibfield
  {journal} {\bibinfo  {journal} {Phys. Rev. D}\ }\textbf {\bibinfo {volume}
  {109}},\ \bibinfo {pages} {124047} (\bibinfo {year} {2024})},\ \Eprint
  {http://arxiv.org/abs/2402.15336} {arXiv:2402.15336 [gr-qc]} \BibitemShut
  {NoStop}%
\bibitem [{\citenamefont {Padmanabhan}\ and\ \citenamefont
  {Padmanabhan}(2017)}]{Padmanabhan:2017qvh}%
  \BibitemOpen
  \bibfield  {author} {\bibinfo {author} {\bibfnamefont {T.}~\bibnamefont
  {Padmanabhan}}\ and\ \bibinfo {author} {\bibfnamefont {H.}~\bibnamefont
  {Padmanabhan}},\ }\href {\doibase 10.1016/j.physletb.2017.07.066} {\bibfield
  {journal} {\bibinfo  {journal} {Phys. Lett. B}\ }\textbf {\bibinfo {volume}
  {773}},\ \bibinfo {pages} {81} (\bibinfo {year} {2017})},\ \Eprint
  {http://arxiv.org/abs/1703.06144} {arXiv:1703.06144 [gr-qc]} \BibitemShut
  {NoStop}%
\bibitem [{\citenamefont {Tichy}\ \emph {et~al.}(2000)\citenamefont {Tichy},
  \citenamefont {Flanagan},\ and\ \citenamefont {Poisson}}]{Tichy:1999pv}%
  \BibitemOpen
  \bibfield  {author} {\bibinfo {author} {\bibfnamefont {W.}~\bibnamefont
  {Tichy}}, \bibinfo {author} {\bibfnamefont {E.~E.}\ \bibnamefont {Flanagan}},
  \ and\ \bibinfo {author} {\bibfnamefont {E.}~\bibnamefont {Poisson}},\ }\href
  {\doibase 10.1103/PhysRevD.61.104015} {\bibfield  {journal} {\bibinfo
  {journal} {Phys. Rev. D}\ }\textbf {\bibinfo {volume} {61}},\ \bibinfo
  {pages} {104015} (\bibinfo {year} {2000})},\ \Eprint
  {http://arxiv.org/abs/gr-qc/9912075} {arXiv:gr-qc/9912075} \BibitemShut
  {NoStop}%
\bibitem [{\citenamefont {Nair}\ \emph {et~al.}(2023)\citenamefont {Nair},
  \citenamefont {Chakraborty},\ and\ \citenamefont {Sarkar}}]{Nair:2022xfm}%
  \BibitemOpen
  \bibfield  {author} {\bibinfo {author} {\bibfnamefont {S.}~\bibnamefont
  {Nair}}, \bibinfo {author} {\bibfnamefont {S.}~\bibnamefont {Chakraborty}}, \
  and\ \bibinfo {author} {\bibfnamefont {S.}~\bibnamefont {Sarkar}},\ }\href
  {\doibase 10.1103/PhysRevD.107.124041} {\bibfield  {journal} {\bibinfo
  {journal} {Phys. Rev. D}\ }\textbf {\bibinfo {volume} {107}},\ \bibinfo
  {pages} {124041} (\bibinfo {year} {2023})},\ \Eprint
  {http://arxiv.org/abs/2208.06235} {arXiv:2208.06235 [gr-qc]} \BibitemShut
  {NoStop}%
\bibitem [{\citenamefont {Ajith}\ \emph {et~al.}(2025)\citenamefont {Ajith}
  \emph {et~al.}}]{Ajith:2024inj}%
  \BibitemOpen
  \bibfield  {author} {\bibinfo {author} {\bibfnamefont {P.}~\bibnamefont
  {Ajith}} \emph {et~al.},\ }\href {\doibase 10.1007/s12036-024-10031-x}
  {\bibfield  {journal} {\bibinfo  {journal} {J. Astrophys. Astron.}\ }\textbf
  {\bibinfo {volume} {46}},\ \bibinfo {pages} {6} (\bibinfo {year} {2025})},\
  \Eprint {http://arxiv.org/abs/2501.04333} {arXiv:2501.04333 [astro-ph.IM]}
  \BibitemShut {NoStop}%
\bibitem [{\citenamefont {Borhanian}\ and\ \citenamefont
  {Sathyaprakash}(2024)}]{Borhanian:2022czq}%
  \BibitemOpen
  \bibfield  {author} {\bibinfo {author} {\bibfnamefont {S.}~\bibnamefont
  {Borhanian}}\ and\ \bibinfo {author} {\bibfnamefont {B.~S.}\ \bibnamefont
  {Sathyaprakash}},\ }\href {\doibase 10.1103/PhysRevD.110.083040} {\bibfield
  {journal} {\bibinfo  {journal} {Phys. Rev. D}\ }\textbf {\bibinfo {volume}
  {110}},\ \bibinfo {pages} {083040} (\bibinfo {year} {2024})},\ \Eprint
  {http://arxiv.org/abs/2202.11048} {arXiv:2202.11048 [gr-qc]} \BibitemShut
  {NoStop}%
\bibitem [{\citenamefont {Maggiore}\ \emph {et~al.}(2020)\citenamefont
  {Maggiore} \emph {et~al.}}]{ET:2019dnz}%
  \BibitemOpen
  \bibfield  {author} {\bibinfo {author} {\bibfnamefont {M.}~\bibnamefont
  {Maggiore}} \emph {et~al.} (\bibinfo {collaboration} {ET}),\ }\href {\doibase
  10.1088/1475-7516/2020/03/050} {\bibfield  {journal} {\bibinfo  {journal}
  {JCAP}\ }\textbf {\bibinfo {volume} {03}},\ \bibinfo {pages} {050} (\bibinfo
  {year} {2020})},\ \Eprint {http://arxiv.org/abs/1912.02622} {arXiv:1912.02622
  [astro-ph.CO]} \BibitemShut {NoStop}%
\bibitem [{\citenamefont {Padmanabhan}(2010)}]{Paddy_book}%
  \BibitemOpen
  \bibfield  {author} {\bibinfo {author} {\bibfnamefont {T.}~\bibnamefont
  {Padmanabhan}},\ }\href@noop {} {\emph {\bibinfo {title} {{\it Gravitation:
  Foundations and frontiers}}}}\ (\bibinfo  {publisher} {Cambridge University
  Press},\ \bibinfo {year} {2010})\BibitemShut {NoStop}%
\bibitem [{\citenamefont {Jacobson}(1995)}]{Jacobson:1995ab}%
  \BibitemOpen
  \bibfield  {author} {\bibinfo {author} {\bibfnamefont {T.}~\bibnamefont
  {Jacobson}},\ }\href {\doibase 10.1103/PhysRevLett.75.1260} {\bibfield
  {journal} {\bibinfo  {journal} {Phys. Rev. Lett.}\ }\textbf {\bibinfo
  {volume} {75}},\ \bibinfo {pages} {1260} (\bibinfo {year} {1995})},\ \Eprint
  {http://arxiv.org/abs/gr-qc/9504004} {arXiv:gr-qc/9504004} \BibitemShut
  {NoStop}%
\bibitem [{\citenamefont {Pesci}(2015)}]{API}%
  \BibitemOpen
  \bibfield  {author} {\bibinfo {author} {\bibfnamefont {A.}~\bibnamefont
  {Pesci}},\ }\href@noop {} {\bibfield  {journal} {\bibinfo  {journal}
  {Entropy}\ }\textbf {\bibinfo {volume} {17}},\ \bibinfo {pages} {5799}
  (\bibinfo {year} {2015})},\ \Eprint {http://arxiv.org/abs/1404.7631}
  {arXiv:1404.7631 [gr-qc]} \BibitemShut {NoStop}%
\end{thebibliography}%

\end{document}